\documentclass[journal]{IEEEtran}
\usepackage{color,amsfonts,colortbl,amsmath,amssymb,booktabs,multirow,psfrag,bm,amsbsy,cite,float}
\usepackage{enumerate}
\usepackage{subfigure}
\ifCLASSINFOpdf

\usepackage[pdftex]{graphicx}
 
\graphicspath{{Figures/}}
\usepackage{epstopdf}
\DeclareGraphicsExtensions{.pdf,.jpeg,.jpg,.png,.eps}\else  
\fi

\begin{document}
\title{A Measurement Based Multilink Shadowing\\ Model for V2V Network Simulations\\ of Highway Scenarios}

\author{\IEEEauthorblockN{Mikael G. Nilsson\IEEEauthorrefmark{1}\IEEEauthorrefmark{2},
		Carl Gustafson\IEEEauthorrefmark{1},
		Taimoor Abbas\IEEEauthorrefmark{2},
		and Fredrik Tufvesson\IEEEauthorrefmark{1}}\\
	
		\IEEEauthorblockA{\IEEEauthorrefmark{1}Lund University, Dept. of Electrical and Information Technology, Box 118, SE-221 00 Lund}, Sweden\\
	
		\IEEEauthorblockA{\IEEEauthorrefmark{2}Volvo Car Corporation, SE-405 31 G\"oteborg, Sweden}\\
	
		email: mikael.nilsson@volvocars.com}


%



\maketitle

\begin{abstract}
Shadowing from vehicles can significantly degrade the performance of vehicle-to-vehicle (V2V) communication in multilink systems, e.g., vehicular \textit{ad-hoc} networks (VANETs). It is thus important to characterize and model the influence of common shadowing objects like cars properly when designing these VANETs. Despite the fact that for multilink systems it is essential to model the joint effects on the different links, the multilink shadowing effects of V2V channels on VANET simulations are not yet well understood. In this paper we present a measurement based analysis of multilink shadowing effects in a V2V communication system with cars as blocking objects. In particular we analyze, characterize and model the large scale fading, both regarding the autocorrelation and the joint multilink cross-correlation process, for communication at 5.9\,GHz between four cars in a highway convoy scenario. The results show that it is essential to separate the instantaneous propagation condition into line-of-sight (LOS) and obstructed LOS (OLOS), by other cars, and then apply an appropriate pathloss model for each of the two cases. The choice of the pathloss model not only influences the autocorrelation but also changes the cross-correlation of the large scale fading process between different links.
By this, we conclude that it is important that VANET simulators should use geometry based models, that distinguish between LOS and OLOS communication. Otherwise, the VANET simulators need to consider the cross-correlation between different communication links to achieve results close to reality.
\end{abstract}


%
\IEEEpeerreviewmaketitle


\section{Introduction}


\IEEEPARstart{E}{very} 
year 1.25 million people are killed in road traffic accidents \cite{WHO}. Two enablers to increase road safety and hopefully reduce the number of fatalities per year are Vehicle-to-Vehicle (V2V) and Vehicle-to-Infrastructure (V2I) communications, which have been introduced, and standardized \cite{IEEE80211}. This V2V and V2I communication will also help to increase traffic efficiency, as well to reduce the environmental impact caused by vehicles.

In intelligent transport systems (ITS) \cite{ETSI-TS-BSA-1,ETSI-TS-BSA-2}, messages are exchanged over the dedicated 5.9\,GHz frequency band, based on the establishment of vehicular ad-hoc networks (VANETs), which can be highly dynamic in nature. The propagation channel at the 5.9\,GHz band is considered to be one of the most critical performance limiting factors for VANETs, due to high penetration losses, high mobility of vehicles, and antennas being relatively close to the ground-level. Many studies have been performed regarding the channel characteristics of the 5.9\,GHz band using channel sounding equipment as well as 802.11p transceivers \cite{Molisch09, Mecklenbrauker11, Cheng07, Boban14, Mangel11-3, Renaudin13}. Based on the channel characteristics and subsequent channel models, simulations of VANETs have been performed by researchers to identify the performance of the network \cite{Abbas15, Boban14, Biddlestone12}. Previous VANET simulation studies have often used rather simplified models. The channel model is a cruicial input to the network simulator, hence it is important that the models used are close to reality. 

V2V is particularly beneficial in situations where the visual line-of-sight (LOS) is obstructed by buildings or other vehicles. However, vehicles and buildings as physical obstructions induce additional propagation losses (shadow fading) by blocking the LOS signals, which in turn reduces the communication range. In such an obstructed-LOS (OLOS) or non-LOS (NLOS) situation scattering and reflections from nearby objects, e.g., buildings, traffic signs, trucks and bridges, enable the signal reception \cite{Abbas11}. In the literature it is reported that a single vehicle on average can induce an additional shadow fading of about 10-20 dB \cite{Boban14b,Vlastaras14} depending upon the shape, size and location of the obstructing vehicle. Moreover a few measurement results regarding obstruction by buildings in different types of street crossings and their corresponding channel models are available in \cite{Mangel11-3,Paschalidis16,Giordano11,Turkka11}. It is important that the model incorporates the effect of buildings and vehicles as obstacles, because ignoring this can lead to an unrealistic assumptions about the performance of the physical layer, both in terms of received signal power as well as interference levels, which in turn can affect the behavior of higher layers in V2V systems. To date, in majority of the findings for V2V communications except \cite{Boban14b,Boban11,Abbas11}, the shadowing impact of vehicles has largely been neglected when modeling the pathloss. To explicitly characterize this impact the following categorization is defined:
\begin{itemize}
	\item Line-of-sight (LOS) is the situation when there is an	optical line-of-sight between the transmitter (TX) and the receiver (RX) antennas.
	\item Obstructed-LOS (OLOS) is the situation when the optical LOS between the TX and RX antennas is obstructed by another vehicle.
	\item Non-LOS (NLOS) is the situation when a building between the TX and RX completely block the LOS (as well	as many other significant multi-path components). This category is typically not found in a highway scenario. 
\end{itemize}
The channel properties for LOS and OLOS are distinct, and their individual analysis is required. No pathloss model, except a geometry based channel model \cite{Boban14} and a measurement based fading model \cite{Abbas15} both published recently, is today available dealing with both the cases in a comprehensive way.

To generate realistic simulations of VANET performance, both a traffic and a network simulator is needed. A common, open source, traffic simulator is SUMO \cite{SUMO2012} which models vehicle positions, exact velocities, inter-vehicle spacings, accelerations, overtaking attitudes, lane-change behaviors, etc. As a network simulator platform many researchers are today using OMNeT++ \cite{OMNeT}, also open source, which is a discrete-event network simulator supporting a variety of static or dynamic routing protocols. Both simulators are then used in the vehicular network simulator, Veins \cite{Veins}. Most reported VANET simulation results have used only LOS channel models as input. One example is the model by Cheng et al. \cite{Cheng07} which is based on outdoor channel sounding at 5.9\,GHz. This model does not classify the measured data into LOS and OLOS, but it represents both small scale fading and the shadowing by the Nakagami-m model. To improve the input to VANET simulation they have in \cite{Abbas15} defined two channel models, one for LOS and one for OLOS. Further improvement is still needed, namely to also include the cross-correlation of the different communication links as input to the VANET simulations. This is important since wireless communication systems using multihop techniques \cite{ETSI-TS-BSA-1,ETSI-TS-BSA-2}, can overcome the issue with shadowed cars in V2V systems. Including the cross-correlation in VANET simulations are performed in \cite{WangX14} and the initial results show a significant impact on the observed protocol performance.

Most of the measurement studies have considered communication between one TX and one RX, i.e., the single link case. Only a few measurement studies have considered communication between one TX and several RX or vice versa, i.e., the multilink case. As pointed out in \cite{Wang08} it is essential to consider the properties of the joint shadowing process in ad hoc peer-to-peer channels. As well, the importance of the correlated link shadowing in mesh networks is shown in \cite{Agrawal09}. An extensive feasibility study regarding correlation models of shadowing are summarized in \cite{Szyszkowicz10} and how surrounding vehicles can change the cross-correlation between two different mobile stations connected to a base station in a small cell configuration are presented in \cite{Maviel13}. Regarding VANETs in \cite{Nilsson15} it is shown that multiple links can be highly correlated and in \cite{Chen16} a first step is taken to model the multilink shadowing versus distance between two receiving cars.

In this paper the multilink shadowing effects are analyzed for a use case where a car is making an emergency brake on the highway and that information is transmitted to the surrounding cars over the air. The purpose of the analysis is to model the joint shadowing process, considering both LOS and OLOS conditions, in a use case such a highway scenario for V2V channels with the help of extensive channel measurement data. The overall goal of the analysis, is also to investigate if previous multilink shadowing models used for cellular communication systems, such as the one presented in \cite{Wang08,Oestges11,Yamamoto06}, can also be applied to VANETs. With this in mind, the main contributions of the paper are: 1) estimation of pathloss model parameters for both LOS and OLOS using maximum likelihood with truncated data. 2) analysis of the autocorrelation and the cross-correlation of the large scale fading processes for concurrent communication links, i.e., the multilink case. 3) a newly developed model for the joint shadowing process, i.e., a cross-correlation model that can be used as input to VANET simulators.

The remainder of the paper is organized as follows. Section~II outlines the V2V measurement test setup, the test scenario, data processing, and pathloss model. The results are presented in section~III, with focus on pathloss model parameter estimation for LOS and OLOS cases separately, auto- and cross-correlation of the shadowing using different pathloss model as input for the large scale fading estimation. To demonstrate the implications of our models, section~IV present some examples of the impact
of the auto- and cross-correlation on the fading durations. Finally, section V concludes the paper.

\begin{figure}[h]
	\centering
	\includegraphics[width=0.48\textwidth]{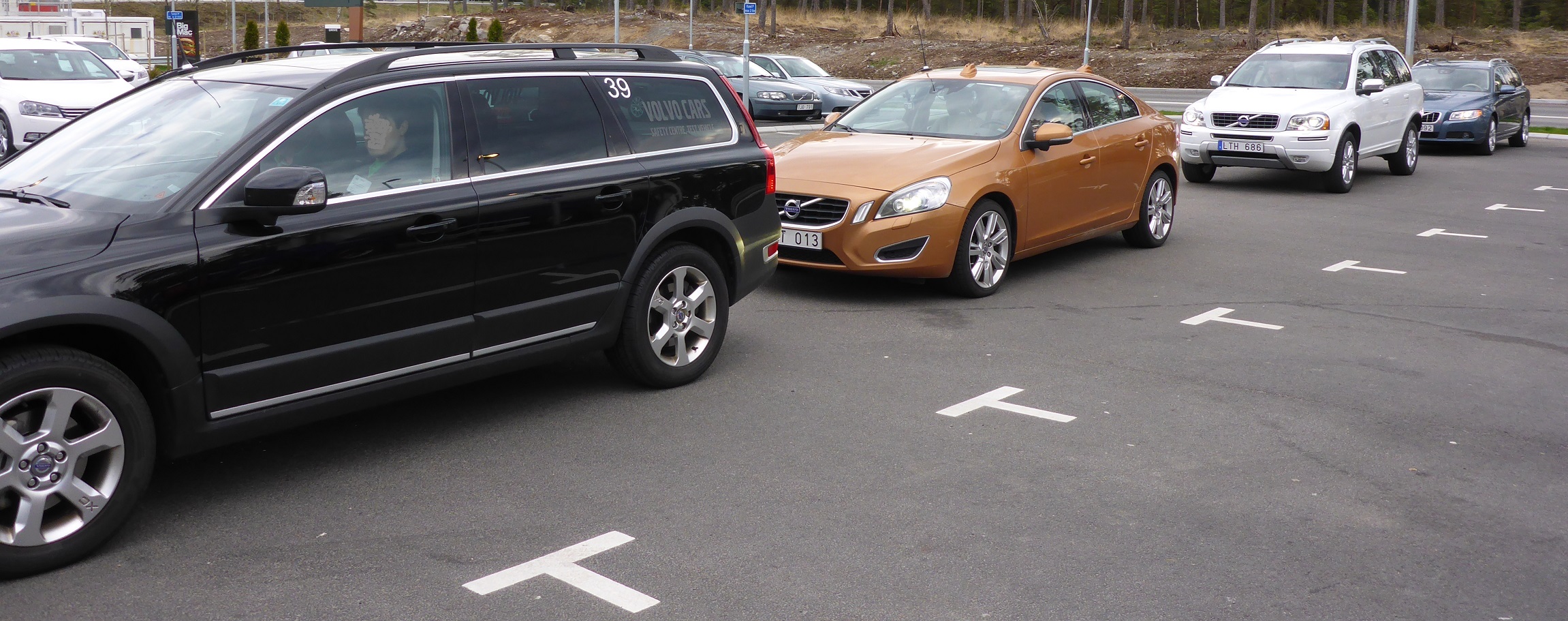}
	\caption{All test cars (Volvo XC70, S60, XC90 and V70, respectively) lined up for the test scenario Convoy A (TS A).}
	\label{fig:Cars}
\end{figure}

\section{Method}
\subsection{Measurement Description}
\label{sec:Meas}
Measurements were performed to capture the multilink behavior between four Volvo cars, i.e., the behavior of in total six simultaneous communication links. The cars, see Fig.~\ref{fig:Cars}, were driving in the same direction on the highway \emph{Rv 40} between Landvetter airport (lat N 57.68014, long  E 12.31441) and Bor\aa s city (lat  N 57.71660, long E 12.91652), Sweden. Each car was equipped with a transceiver from Kapsch TrafficCom AB \cite{EVK3300}, using the communication standard IEEE 802.11p \cite{IEEE80211}. The antennas used in the test have been developed specifically for V2V communication at 5.9\,GHz. Videos and GPS coordinates were recorded from each vehicle. Several measurement scenarios were tested and for each scenario there were simultaneous LOS and OLOS links. 

\begin{table}[h]
	\renewcommand{\arraystretch}{1.3}
	\caption{Measurement Parameters}
	\label{tab:Setup}
	\centering
	\begin{tabular}{l l}
		\hline 
		{\bfseries Parameter} & {\bfseries Value(s)}\\ 
		\hline 
		Standard & IEEE 802.11p\\
		Center frequency & 5.9\,GHz \\
		Data rate & 6\,Mbit/s \\ 
		Packet rate & 10\,Hz \\ 
		Packet size & 100, 500 and 1500 bytes every 100\,ms. \\
		TX output power & +23\,dBm, limit in \cite{ECC15}, +23\,dBm/MHz e.i.r.p \\
		RX sensitivity & -97\,dBm @ Packet Error Rate of 10\,\%\\ 
		S60 (mid size) & 1 antenna at roof, height 145\,cm, cable loss, 1.0\,dB.\\
		& 1 at front windscreen and 1 at rear windscreen,\\
		& both with the height 135\,cm and cable loss 1.0\,dB.\\ 
		V70 (large size)  & 1 antenna at roof, height 155\,cm. Cable loss, 3.5\,dB.\\  
		XC70 (large size) & 1 antenna at roof, height 160\,cm. Cable loss, 3.5\,dB.\\
		XC90 (large SUV) & 1 antenna at roof, height 178\,cm. Cable loss, 3.5\,dB.\\  
		\hline 
	\end{tabular} 
\end{table}

\begin{figure*}[t!]
	\centering
	\includegraphics[trim = 10mm 8mm 10mm 7mm, clip=true,width=0.99\textwidth]{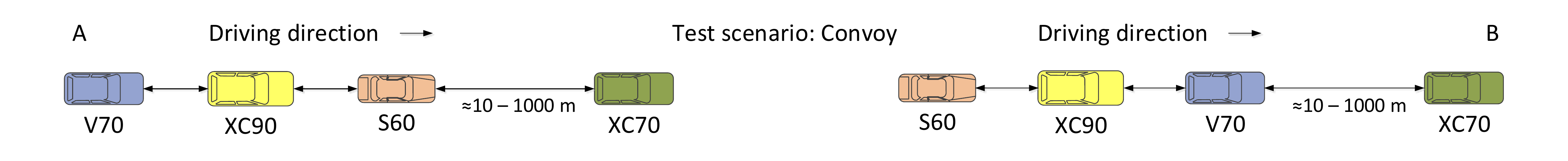}
	\caption{Test scenarios, Convoy A (TS A) and Convoy B (TS B). Volvo S60 is a compact sedan, Volvo V70 is a large station wagon, XC70 same as V70 but taller, and XC90 is a Sport Utility Vehicle (SUV). The distance between the first and second car was intentionally changed between 10\,m to 1000\,m. The distances between the other cars was unintentionally changed, it was the traffic situation that made these decisions. Distances are not correctly scaled.}
	\label{fig:Meas_scenarios}
\end{figure*}


\subsection{Measurement Setup}
\label{sec:Setup}
\begin{figure}[b]
	\centering
	\includegraphics[trim = 20mm 23mm 20mm 5mm, clip, width=0.48\textwidth]{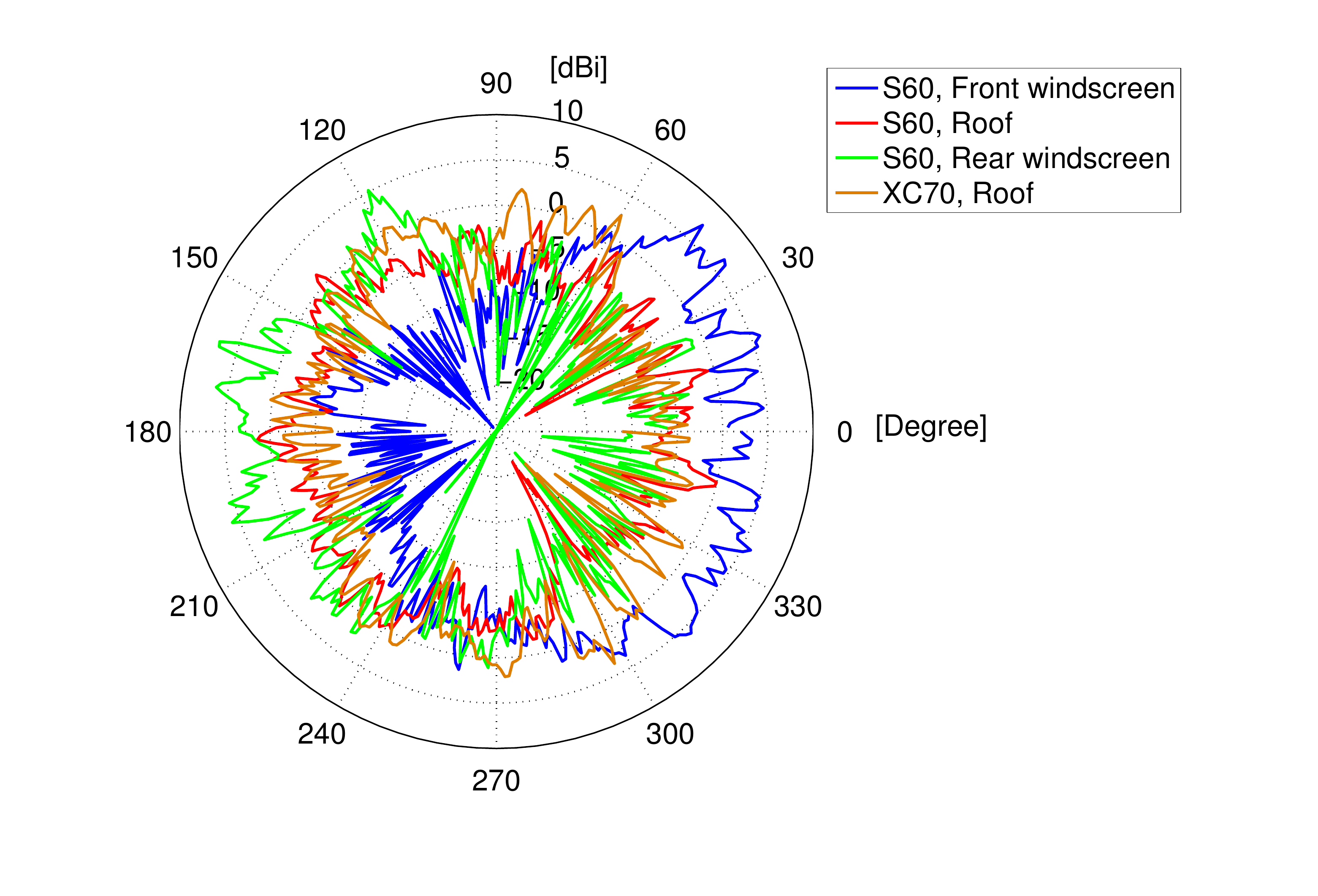}
	\caption{Measured antenna pattern (vertical polarization), mounted on the Volvo S60 and the Volvo XC70. The front of the cars is at $0\,^{\circ}$. The antenna pattern of the V70 and XC90 are similar to the presented XC70 since all three models were using the same antenna type and the design of the roof area is similar.}
	\label{fig:S60_antenna_pattern}
\end{figure}

Each transceiver was connected via a coaxial cable to an antenna mounted on the roof of each car, i.e., a Volvo V70 (station wagon), a XC70 (taller station wagon) and a XC90 (SUV - Sport Utility Vehicle). Inside a Volvo S60 (compact sedan) two transceivers were installed, the first one connected to an antenna mounted on the roof whereas the second one, which had antenna diversity capability, was connected to two antenna elements mounted inside the car at the front and rear windscreens. The second transceiver make it possible to analyze the performance of hidden antennas and antenna diversity. All the antenna elements were vertically polarized and approximately omni-directional in the azimuthal plane, but the antenna pattern after mounting is certainly not, as seen in Fig. \ref{fig:S60_antenna_pattern}. During the measurements one transceiver transmitted packets, bursts of pseudo random data with size and rate according to Table \ref{tab:Setup}. Simultaneously all the other transceivers were recording received packets. Next, another transceiver transmitted packets and all other were recording. The hopping sequence, with the 10\,Hz rate, between TX and RX states is defined in the \cite{IEEE80211}.

\subsection{Scenario Description}
\label{sec:Scenario}
While many measurements were performed under varying traffic and road conditions, one major scenario is presented and analyzed in this paper, namely a convoy scenario with two different configurations, see Fig.~\ref{fig:Meas_scenarios}. The measurements were performed with a typical average velocity of 25\,m/s (90\,km/h). The convoy test scenarios (TS), TS A and TS B, were measured with different positions of the cars to study obstruction losses due to different car sizes. The highway, \emph{Rv 40}, is characterized by low traffic in non-rush hours, guard rails in the middle of the road, some bridges, other structures, and with forest besides the road. The measurements are expected to characterize a normal to rich scattering environment.

\subsection{Data processing}
Each transceiver transmitted packets with a 10 Hz repetition rate and the transceivers reported received signal strength (RSSI) in dB with a 1 dB resolution for successfully decoded packets. All transceivers were synchronized to GPS, meaning that they had a common time reference so that the time instants where the transmission failed and the number of missing packets could be identified. In addition, the video recordings were synchronized with the transceiver log files by using the available geometric information in the videos, e.g., a bridge was used as the reference point for our measurements. By this, the videos have served as the foundation for the LOS and OLOS separation process. We define the OLOS case as when the camera in the RX car can see less then half of the TX car at the ground level.

The knowledge of the missing packets is used in the maximum likelihood (ML) based estimations of the pathloss parameters to compensate for the influence of the sensitivity level of the receiver, often due too long distances between TX and RX. The ordinary least square (OLS) estimation method does not consider this issue, i.e., when only successfully decoded packets are used in the estimation. As mentioned in \cite{Abbas14,Kokalj-Filipovic15,Gustafson15} it is important to use the information regarding distances between TX and RX also for the lost packets when estimating pathloss parameters for LOS and OLOS channel models. In this paper we are using the ML estimation method described in \cite{Gustafson15}. The censoring due to lost packets does not occur at a fixed level. Instead, there is a soft censoring that occurs for data below RSSI values of -94\,dBm (which is close to the sensitivity limit of the receivers specified by the manufacturer). For this reason, all data below -94\,dBm are regarded as censored data in the analysis. Seldom a packet can be lost also at short distances due to destructive addition of different multipath components or interference. These lost packets are also considered in the estimation algorithm using the ML method.

To be able to estimate the pathloss parameters, all RSSI values of the communication link are compensated with respect to the TX output power, +23\,dBm, and with the cable losses in respective car, see Table~\ref{tab:Setup}. We call this resulting value the channel gain, more specifcally, \textit{channel gain = RSSI - TX\,output\,power + TX\,cable\,loss + RX\,cable\,loss}.

It should be noted that both the absolute values of the simultaneous signal strengths and the joint distribution of the shadow fading process and its autocorrelation function are of interest for channel characterization and performance evaluations. For this specific measured scenario, of course the received signal strengths and the estimated packet error rates can be used directly, but for a realistic channel model with arbitrary distances between the vehicles one has to separate the correlated large scale fading process and the distance dependent pathloss. The large scale fading is estimated by subtraction of the distance dependent pathloss from the channel gain value. Typically when estimating the large scale fading it is common to perform averaging of the received power over $10\lambda$ or more to reduce the effect of the small scale fading. Since our transceivers transmitted with a transmission rate of 10\,Hz according to \cite{IEEE80211} we received a packet approximately every $50\lambda$ at 25~m/s. Every RSSI value is influenced by the relative distance between TX and RX, as well the small and large scale fading effects. For the best trade off between reducing the small scale fading and the sampling rate in time, we collected the channel gain values in 0.4 second bins and calculated the average channel gain for each bin. The influence of obstruction by other cars is considered as a random time-correlated part of the large scale fading process. 



\subsection{Pathloss Model}
As pointed out in \cite{Boban14,Abbas15} it is crucial to distinguish between LOS and OLOS communication, by using different pathloss models for LOS and OLOS, when performing VANET simulations to get results closer to reality. A classical pathloss model for both LOS and OLOS is the log-distance power law model with a single slope. In units of dB, the single slope model can be written as 
\begin{eqnarray}
PL(d)=PL(d_{0})+10\alpha\mathrm{log}_{10}\left(\frac{d}{d_{0}}\right)+\Psi_{\sigma_{\mathrm{single}}}, \ \ d\geq d_{0}, 
\label{eq:One_slope}
\end{eqnarray}
where $d$ is the distance, $\alpha$ is the pathloss exponent, $PL(d_{0})$ is the pathloss at a reference distance of $d_{0}$=10\,m and $\Psi_{\sigma_{\mathrm{single}}}$ is a random variable that describes the large scale fading about the distance-dependent mean pathloss. From our measurements the single slope fits well to the OLOS case, see Fig.~\ref{fig:OLOS_model}. A dual slope model \cite{Cheng07} have been investigated. However, due to the low number of samples at larger distances for some of the OLOS communication links, the estimated pathloss model will not be physically correct. For the LOS case the data has a clear two-ray behavior, as seen in Fig.~\ref{fig:LOS_model}. 
\begin{figure}[h!]
	\centering
	\includegraphics[trim = 10mm 10mm 10mm 10mm, clip, width=0.48\textwidth]{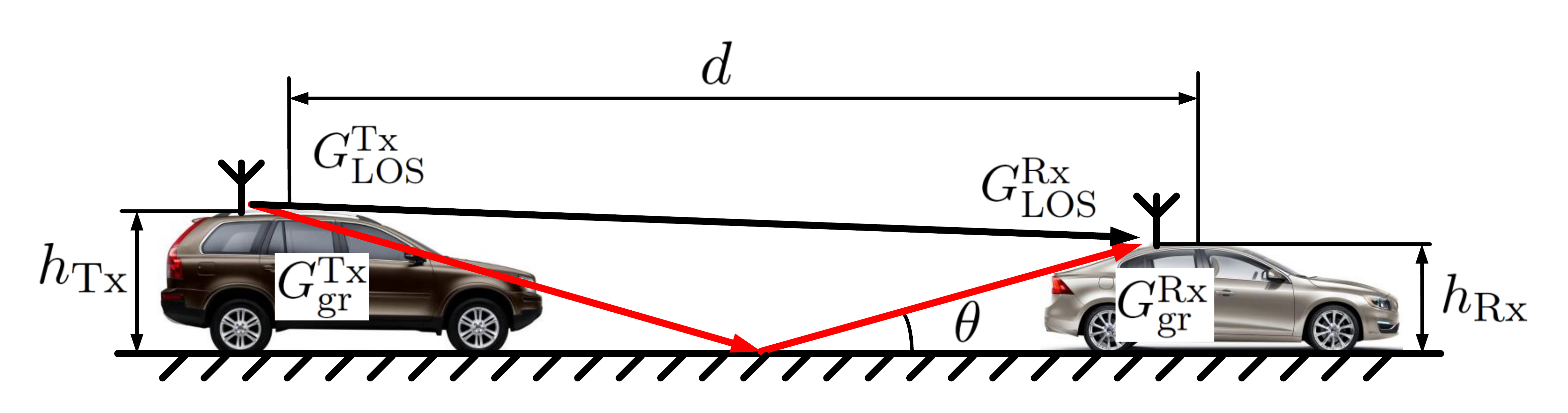}
	\caption{The two-ray model.}
	\label{fig:Tworay_model}
\end{figure}
While previous work has concluded that LOS links can be modeled well with a two-ray ground reflection model \cite{Liu16}, we provide further analysis to account for differences in antenna patterns \cite{Goldsmith05}, car bodies, and antenna patterns. The general two-ray pathloss model is based on the simple geometrical setup shown in Fig.~\ref{fig:Tworay_model}, and is given by\footnote{We note that this is for a narrowband case, where the system bandwidth is such that it is not possible to resolve the LOS from the ground reflection.} 


\begin{align*}
	&PL(d)= 20\mathrm{log}_{10}\left(\frac{4\pi}{\lambda}\right)+ \Psi_{\sigma_{_{\mathrm{tr}}}}\\&-20\mathrm{log}_{10}\left| \sqrt{g_{\mathrm{LOS}}}\mathrm{e}^{\mathrm{j}\phi_{\mathrm{LOS}}} \ \frac{\mathrm{e}^{-{\mathrm{j}}k d_{\mathrm{LOS}}}}{d_{\mathrm{LOS}}}+\sqrt{g_{\mathrm{gr}}}\mathrm{e}^{\mathrm{j}\phi_{\mathrm{gr}}} \ \Gamma\frac{\mathrm{e}^{-{\mathrm{j}}k d_{\mathrm{gr}}}}{d_{\mathrm{gr}}}\right|,    
\end{align*}
where $\sqrt{g}\mathrm{e}^{{\mathrm{j}\phi}}$ is the combined effect of the complex amplitude gains of the TX and RX antennas in the directions of the LOS and the ground reflection components, respectively, such that
\begin{align}
	\sqrt{g_\mathrm{LOS}}\mathrm{e}^{\mathrm{j}\phi_{LOS}}&=\sqrt{G^{\mathrm{Tx}}_{\mathrm{LOS}}G^{\mathrm{Rx}}_{\mathrm{LOS}}}\mathrm{e}^{\mathrm{j}(\phi^{\mathrm{Tx}}_{\mathrm{LOS}}+\phi^{\mathrm{Rx}}_{\mathrm{LOS}})},\label{eq:ant1} \\
		\sqrt{g_\mathrm{gr}}\mathrm{e}^{\mathrm{j}\phi_{gr}}&=\sqrt{G^{\mathrm{Tx}}_{\mathrm{gr}}G^{\mathrm{Rx}}_{\mathrm{gr}}}\mathrm{e}^{\mathrm{j}(\phi^{\mathrm{Tx}}_{\mathrm{gr}}+\phi^{\mathrm{Rx}}_{\mathrm{gr}})}. \label{eq:ant2}
\end{align}

Ideally, the variables in (\ref{eq:ant1}) and (\ref{eq:ant2}) should be included in the estimation based on accurate 3-D measurement data of the complex antenna response for each antenna. However, as we currently do not have access to such measurements, nor the angle of arrival or departure. We instead use a modified version of this model, where we only use fixed average values of the combined antenna gains for the LOS and ground reflection components, respectively. These are denoted $\bar{g}_{\mathrm{LOS}}$ and $\bar{g}_{\mathrm{gr}}$. By factoring out the term for the combined antenna response of the LOS component, with $\Delta\phi=\phi_{\mathrm{gr}}-\phi_{\mathrm{LOS}}$, we can rewrite the two-ray pathloss model as 

\begin{align}
	\begin{split}
		&PL(d)= 20\mathrm{log}_{10}\left(\frac{4\pi}{\lambda}\right)-10\mathrm{log}_{10}(\bar{g}_{\mathrm{LOS}}) + \Psi_{\sigma_{_{\mathrm{tr}}}}\\&-20\mathrm{log}_{10}\left| \frac{\mathrm{e}^{-{\mathrm{j}}k d_{\mathrm{LOS}}}}{d_{\mathrm{LOS}}}+\sqrt{\frac{\bar{g}_{\mathrm{gr}}}{\bar{g}_{\mathrm{LOS}}}}\mathrm{e}^{\mathrm{j}\Delta\phi} \ \Gamma\frac{\mathrm{e}^{-{\mathrm{j}}k d_{\mathrm{gr}}}}{d_{\mathrm{gr}}}\right|.   
		\label{eq:Two_ray3}
	\end{split}
\end{align} 
 The parameter $k=2\pi/\lambda$ is the wavenumber, and the propagation distances for the LOS and ground reflection components are given by
\begin{eqnarray}
	d_{\mathrm{LOS}}=\sqrt{d^2+(h_{\mathrm{Tx}}-h_{\mathrm{Rx}})^2}, \\
	d_{\mathrm{gr}}=\sqrt{d^2+(h_{\mathrm{Tx}}+h_{\mathrm{Rx}})^2}. 
\end{eqnarray}
The effective ground reflection coefficient is then given by
\begin{align}
\begin{split}
	\Gamma_{v}&=\frac{\epsilon_{r}\mathrm{sin} \ \theta-\sqrt{\epsilon_{r}-\mathrm{cos^{2} \ \theta}}}{\epsilon_{r}\mathrm{sin}\ \theta+\sqrt{\epsilon_{r}-\mathrm{cos^{2} \ \theta}}}, \\ 
	\Gamma_{h}&=\frac{\mathrm{sin} \ \theta-\sqrt{\epsilon_{r}-\mathrm{cos^{2} \ \theta}}}{\mathrm{sin} \ \theta+\sqrt{\epsilon_{r}-\mathrm{cos^{2} \ \theta}}}, \label{eq:reflection}
	\end{split}
\end{align}
for vertical and horizontal polarization, respectively, where
\begin{align}
	\mathrm{sin} \ \theta&=(h_{\mathrm{Tx}}+h_{\mathrm{Rx}})/d_{\mathrm{gr}}, \\
	\mathrm{cos} \ \theta&=d/d_{\mathrm{gr}}.
\end{align}

In (\ref{eq:reflection}), $\epsilon_{r}$ is the relative effective complex permittivity of the ground. We here assume that the electrical properties of this ground reflection can be approximated by a single fixed effective permittivity value, which we model as $\epsilon_{r}=\epsilon'-\mathrm{j}\epsilon''=5-j0.2$, based on the results in \cite{Jaselskis03}. Here, we note that the estimation results were not affected much when considering different possible values of $\epsilon_{r}$ within the ranges presented in \cite{Jaselskis03}. The large scale fading of the two-ray model is for simplicity also assumed to be normal. 

\begin{figure}[h]
	\centering
	\includegraphics[width=0.45\textwidth]{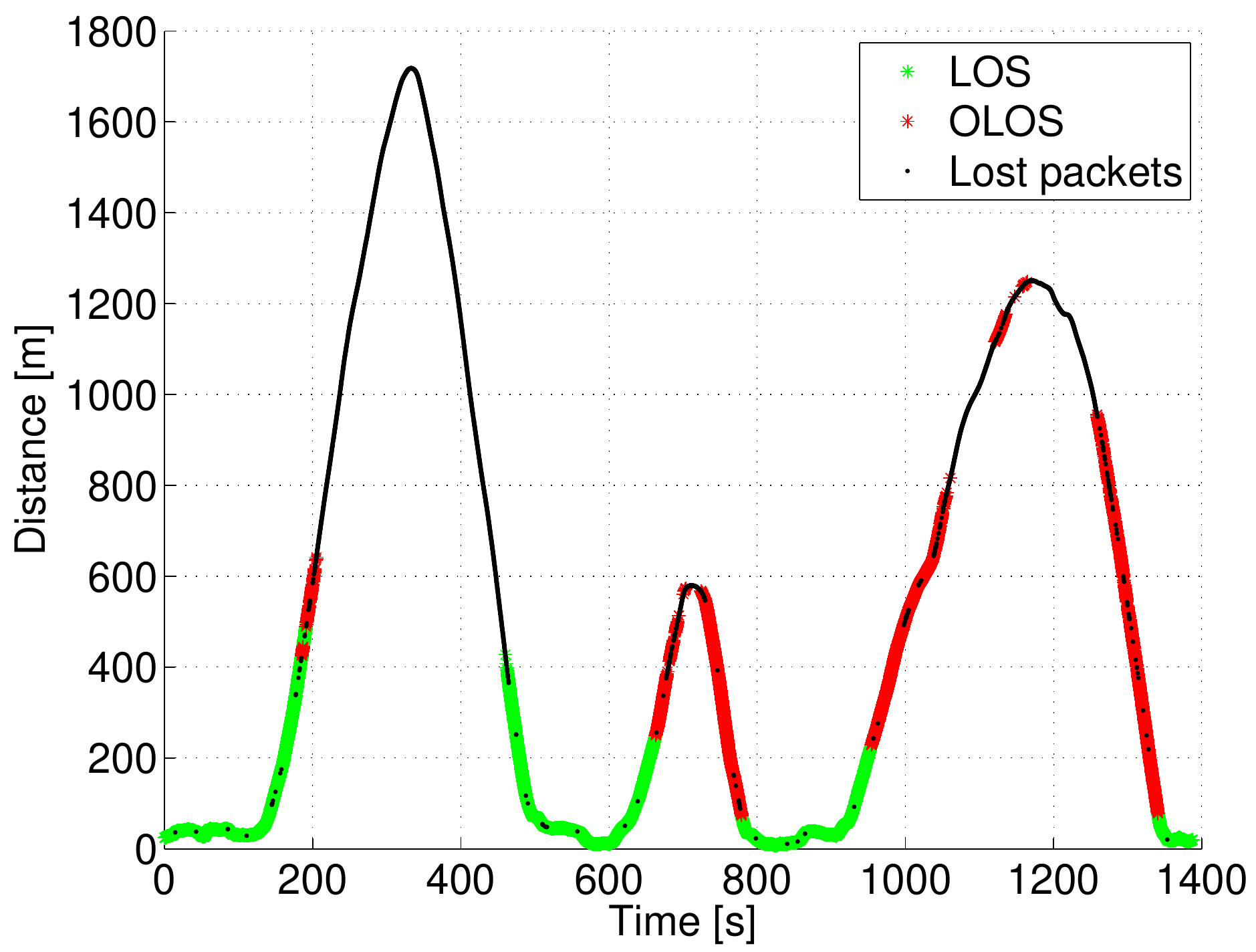}
	\caption{Measured distance between TX and RX car of communication link, XC70-S60M, in TS A.}
	\label{fig:DISTvsTIME}
\end{figure}

\section{Result and discussion}
\subsection{Pathloss Model Estimation}
In total 18 communication links have been investigated and the estimated parameters for the two-ray model (LOS case) and the single slope model (OLOS case) are found in Table~\ref{tab:PL_para}. In TS A six links, all using the roof antennas, are presented and in TS B twelve links are presented. In TS B all three antennas on the S60 are presented, front windscreen (S60F), roof (middle) antenna (S60M), and rear windscreen (S60R). The number of data samples used for the estimation is $m$, and out of these  samples, $m_{c}$ is the number of censored samples which are also used for the estimation. The minimum and maximum distance ($\leq$1000\,m) between TX and RX are, $d_{min}$ and $d_{max}$ respectively. As mentioned before regarding the two-ray model, $\bar{g}_{\mathrm{LOS}}$ and $\bar{g}_{\mathrm{gr}}$ are the combined antenna gains for the TX and RX antennas in the direction of the LOS and ground reflection components, respectively, and, $\Delta\phi$ is the combined (TX and RX) antenna phase response between the LOS and ground reflection component. For the LOS case, the following two-ray parameters of (\ref{eq:Two_ray3}) are being estimated using the ML method: $\bar{g}_{\mathrm{LOS}}$, ${\bar{g}_{\mathrm{gr}}/\bar{g}_{\mathrm{LOS}}}$, $\Delta\phi$ and the large scale fading standard deviation, $\sigma$. For the OLOS case, the following parameters in (\ref{eq:One_slope}) are estimated using ordinary least squares: the pathloss at a reference distance of 10\,m, $PL(d_{0})$, the pathloss exponent, $\alpha$, and the large scale fading standard deviation, $\sigma$.

\begin{figure}[t]
	\centering
	\includegraphics[width=0.49\textwidth]{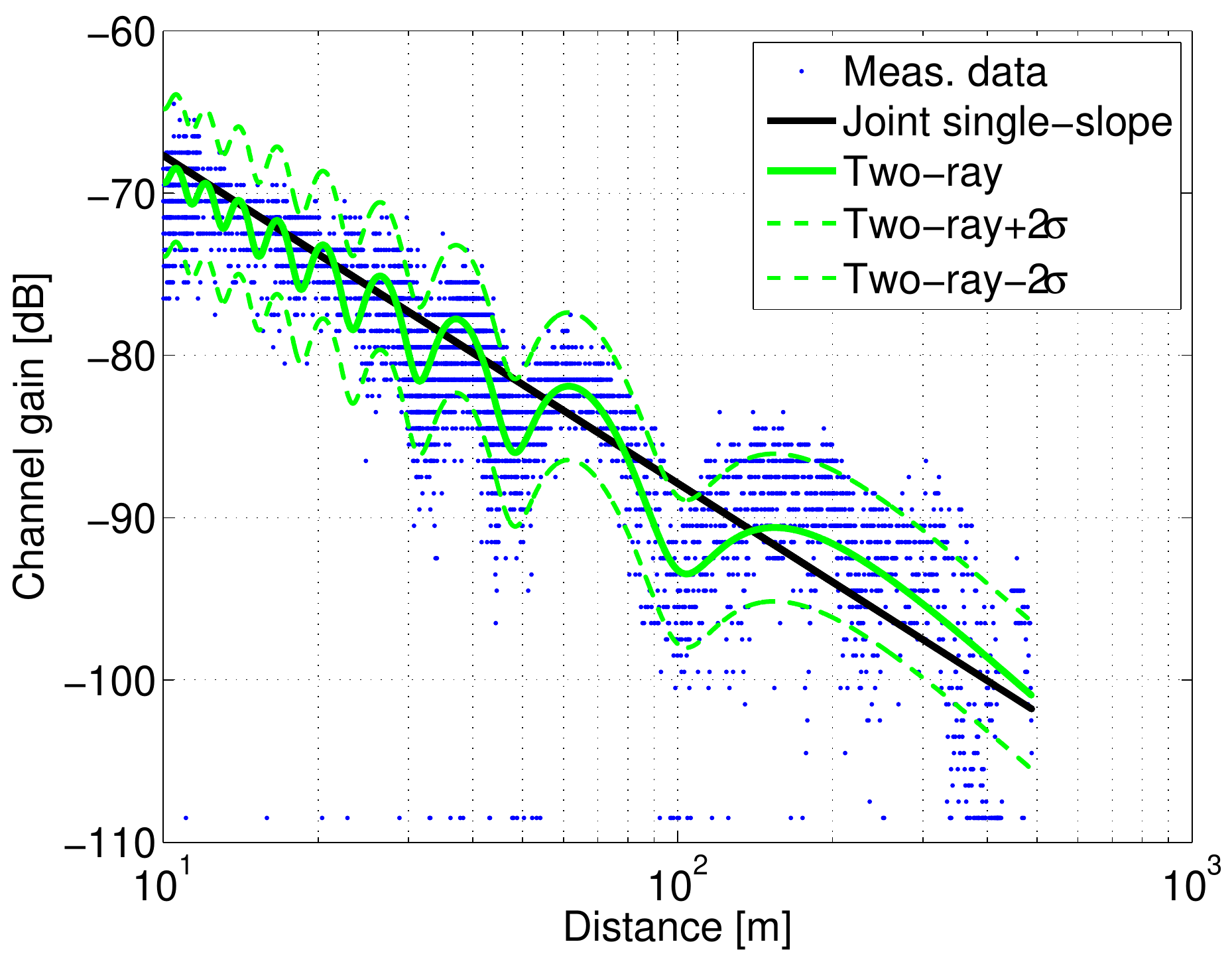}
	\caption{ Measured LOS channel gain of the communication XC70-S60M in TS A and the two-ray model, as well the joint single slope model.}
	\label{fig:LOS_model}
\end{figure}

\begin{figure}[h]
	\centering
	\includegraphics[width=0.49\textwidth]{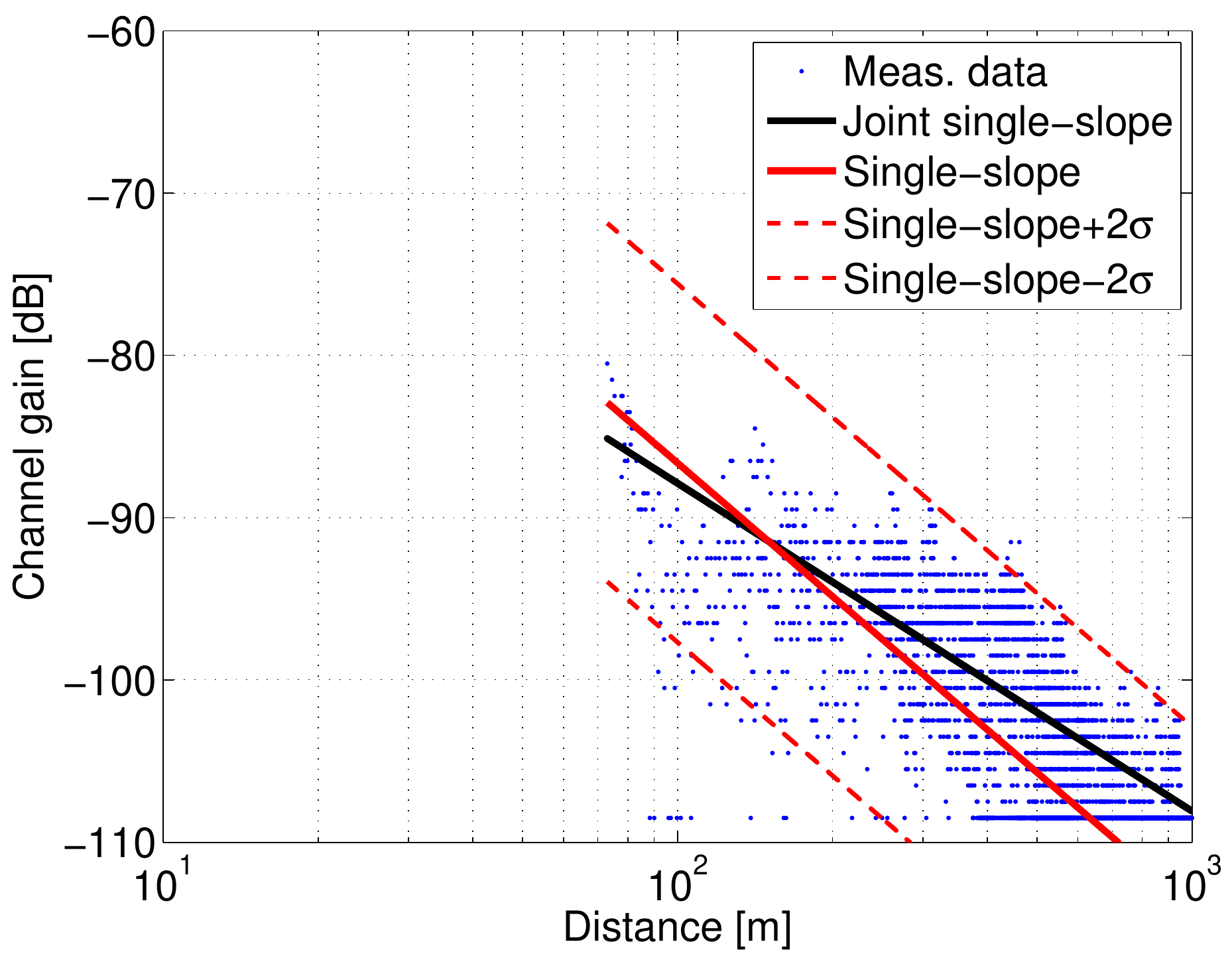}
	\caption{Measured OLOS channel gain of the communication XC70-S60M in TS A and the single slope model, as well the joint single slope model.}
	\label{fig:OLOS_model}
\end{figure}
Here, we note that the uncertainty (more specifically, the standard error) of the estimated parameters depend on the number of samples, the measurement sample distances, the amount of censored samples as well as the pathloss model and its parameters \cite{Gustafson15b}. Furthermore, the effective number of samples are likely reduced due to correlated large scale fading samples that are measured closely in time. The standard errors are different for the two-ray model and the single slope pathloss model. For these reasons, we stress the importance of having a large sample size, and sample distances that are rather evenly distributed, to achieve accurate estimates. In this paper, we have not derived the standard errors of the estimates for each measurement link, but leave that for future work. Instead, as a rule of thumb, we require that $d_{max}/d_{min}\geq10$, the data samples to some extent shall be evenly distributed in between $d_{min}$ and $d_{max}$ with no big gaps and that the number of samples is greater than 1000. Estimated values in black color in Table~\ref{tab:PL_para} fulfill these input requirements and estimated values in red do not. As an example, the communication from XC70 to S60 in TS A fulfill these criteria. In Fig. \ref{fig:DISTvsTIME} the distances between the two cars are shown as a function of time. The measured channel gain and estimated pathloss model for LOS and OLOS data is shown as a function of distance in Fig.~\ref{fig:LOS_model} and Fig.~\ref{fig:OLOS_model}, respectively. Here, joint single slope refers to the pathloss model using the ML estimation method that was previously presented in \cite{Nilsson15} using the same measurement data, that was based on all communications links from both TS A and TS B, where as the model did not distinguish between LOS and OLOS. 
%
%
\begin{table*}[t]
	\renewcommand{\arraystretch}{1.3}
	\caption{Estimated pathloss parameters of a two-ray model for LOS and a single slope model for OLOS.}
	\label{tab:PL_para}
	\centering
	\begin{tabular}{|c|c c c c c c c c|c c c c c c c|}
		\hline
		\textbf{Link}  & \multicolumn{8}{c|}{\bfseries LOS: Two-ray model} & \multicolumn{7}{c|}{\bfseries OLOS: single slope model} \\
		TX-RX & $m$ & $m_{c}$ & $d_{min}$ & $d_{max}$ & $\bar{g}_{\mathrm{LOS}}$ & $\frac{\bar{g}_{gr}}{\bar{g}_{LOS}}$ & $\Delta_{\phi}$ & $\sigma$ & $m$ & $m_{c}$ & $d_{min}$ & $d_{max}$ & $PL(d_{0})$ & $\alpha$ & $\sigma$ \\
		&  &  & [m] & [m] & [dB] & [dB] & [deg] & [dB] &  &  & [m] & [m] & [dB] & [dB] & \\  
		\hline\hline
		
		\textbf{TS A} & \multicolumn{8}{c|}{} & \multicolumn{7}{c|}{}\\ \hline
		
		XC70-S60M & 5759 & 70 & 8 & 488 & -0.8 & -6.42 & -34.53 & 3.12 & 4126 & 1858 & 73 & 1000 & 59.53 & 2.73 & 5.52 \\ \hline
		
		XC70-XC90 & 2633 & 143 & 36 & 721 & -0.98 & -4.60 & -18.74 & 3.02  & 6614 & 1826 & 32 & 1000 & 71.32 & 1.90 & 4.12 \\ \hline
		
		XC70-V70 & \color{red} 775 & \color{red} 2 & \color{red} 83 & \color{red} 321 & \color{red} 0.65 & \color{red} 4.36 & \color{red} -9.65 & \color{red} 3.32  & 8207 & 1580 & 74 & 998 & 65.19 & 2.04 & 5.20 \\ \hline
		
		S60M-XC90 & 12236 & 56 & 12 & 294 & 2.66 & -3.33 & 7.61 & 3.17 & \color{red} 856 & \color{red} 26 & \color{red} 42 & \color{red} 291 & \color{red} 70.72 & \color{red} 1.63 & \color{red} 4.89 \\ \hline
		
		S60M-V70 & \color{red} 3419 & \color{red} 7 & \color{red} 59 & \color{red} 372 & \color{red} 6.22 & \color{red} 2.30 & \color{red} -30.13 & \color{red} 3.59 & \color{red} 9720 & \color{red} 108 & \color{red} 54 & \color{red} 372 & \color{red} 68.63 & \color{red} 1.35 & \color{red} 4.82 \\ \hline
		
		XC90-V70 & \color{red} 12891 & \color{red} 90 & \color{red} 37 & \color{red} 116 & \color{red} 1.58 & \color{red} -8.02 & \color{red} -4.82 & \color{red} 1.71 & \color{red} 238 & \color{red} 1 & \color{red} 38 & \color{red} 99 & \color{red} 68.73 & \color{red} 1.73 & \color{red} 1.73 \\ \hline
		
		\textbf{TS B} & \multicolumn{8}{c|}{} & \multicolumn{7}{c|}{}\\ \hline
		
		XC70-V70 & 6972 & 54 & 11 & 547 & 6.42 & -8.12 & -11.72 & 2.80 & \color{red} 5284 & \color{red} 2455 & \color{red} 217 & \color{red} 943 & \color{red} 29.62 & \color{red} 4.18 & \color{red} 6.58 \\ \hline
		
		XC70-XC90 & \color{red} 1230 & \color{red} 61 & \color{red} 99 & \color{red} 578 & \color{red} 0.27 & \color{red} -3.86 & \color{red} -10.00 & \color{red} 3.51 & 10400 & 3348 & 35 & 998 & 69.82 & 2.02 & 4.43 \\ \hline
		
		XC70-S60M & \color{red} 356 & \color{red} 24 & \color{red} 81 & \color{red} 464 & \color{red} -3.13 & \color{red} -5.80 & \color{red} 1.83 & \color{red} 3.05 & 5891 & 2185 & 74 & 922 & 68.29 & 2.30 & 5.61 \\ \hline
		
		V70-XC90 & \color{red} 12216 & \color{red} 58 & \color{red} 20 & \color{red} 165 & \color{red} 0.07 & \color{red} -6.97 & \color{red} -11.83 & \color{red} 2.57 & \color{red} 870 & \color{red} 21 & \color{red} 64 & \color{red} 166 & \color{red} 76.13 & \color{red} 1.44 & \color{red} 4.46 \\ \hline
		
		V70-S60M & \color{red} 1690 & \color{red} 42 & \color{red} 37 & \color{red} 116 & \color{red} -4.83 & \color{red} -13.01 & \color{red} 344.39 & \color{red} 3.71 & \color{red} 5363 & \color{red} 161 & \color{red} 51 & \color{red} 214 & \color{red} 86.88 & \color{red} 0.54 & \color{red} 4.77 \\ \hline
		
		XC90-S60M & \color{red} 7063 & \color{red} 48 & \color{red} 11 & \color{red} 54 & \color{red} -5.30 & \color{red} -9.40 & \color{red} -83.36 & \color{red} 2.28 & N/A & N/A & N/A & N/A & N/A & N/A & N/A \\ \hline
		
		
		XC70-S60F & \color{red} 359 & \color{red} 9 & \color{red} 79 & \color{red} 463 & \color{red} 3.58 & \color{red} 3.13 & \color{red} 22.17 & \color{red} 3.22 & 5913 & 1855 & 72 & 923 & 60.84 & 2.67 & 6.13 \\ \hline
		
		XC70-S60R & \color{red} 359 & \color{red} 12 & \color{red} 79 & \color{red} 460 & \color{red} -6.59 & \color{red} -4.02 & \color{red} 0.07 & \color{red} 2.34 & 5913 & 1998 & 72 & 920 & 82.54 & 1.62 &  3.30 \\ \hline
		
		V70-S60F & \color{red} 1697 & \color{red} 13 & \color{red} 34 & \color{red} 114 & \color{red} -12.1 & \color{red} 65.82 & \color{red} 325.43 & \color{red} 4.76 & \color{red} 5376 & \color{red} 31 & \color{red} 50 & \color{red} 213 & \color{red} 80.96 & \color{red} 0.72 & \color{red} 4.40 \\ \hline
		
		V70-S60R & \color{red} 1697 & \color{red} 15 & \color{red} 34 & \color{red} 114 & \color{red} -12.1 & \color{red} -20.19 & \color{red} 359.22 & \color{red} 2.84 & \color{red} 5376 & \color{red} 130 & \color{red} 50 & \color{red} 213 & \color{red} 87.15 & \color{red} 1.16 & \color{red} 3.19 \\ \hline
		
		XC90-S60F & \color{red} 7084 & \color{red} 30 & \color{red} 10 & \color{red} 53 & \color{red} -1.74 & \color{red} -13.25 & \color{red} 23.64 & \color{red} 2.66 & N/A & N/A & N/A & N/A & N/A & N/A & N/A \\ \hline
		
		XC90-S60R & \color{red}7084 & \color{red}29 & \color{red}10 & \color{red}53 & \color{red}-11.7 & \color{red} -13.55 & \color{red}169.91 & \color{red} 2.80 & N/A & N/A & N/A & N/A & N/A & N/A & N/A \\ \hline
		
	\end{tabular}
\end{table*}
\subsection{Shadow Fading Autocorrelation}
The cars in the VANET will shadow each other from time to time. Once a car is shadowed, it will be so for some time interval or corresponding traveled distance interval. In general one could say that, with correlated links, if one packet fails there is a higher probability that the next packet will fail as well , which affects the performance of relaying and repetition schemes. The shadow fading autocorrelation is in general a function of time and space. However, estimating the autocorrelation as a function of both the time and the exact locations at both the TX and RX requires many more points than what is typically obtained during field measurements \cite{Szyszkowicz10}. Instead, we follow a more feasible approach of modeling the shadow fading autocorrelation as a function of distance. This is a common approach in V2V channel characterization and also has the benefit of making it easier to implement V2V channels in network simulators. Using the measured instantaneous velocities of each car, we have calculated the absolute distances traveled for each car. Since both the RX and the TX cars are moving, i.e., both ends of the communication link are moving, the traveled distance is $d_i=\bar{v_i}t$, where $\bar{v_i}$ is the average of the instant velocities of the RX and the TX cars and $t$ is 0.4\,s, i.e., the averaging bin size to reduce small scale fading effects. For a given link, the distance traveled during each recorded measurement sample is gathered in the vector \textbf{d}, with elements $d_{1} < d_{2} < d_{3} < \ldots < d_{N}$. These distances are irregularly sampled, and therefore we estimate the sample autocorrelation based on distance bins, $\Delta d_{bin}$. The size of the distance bins, $\Delta d_{bin}$, are calculated as the average velocity during the test scenario for the specific link times 0.4\,s. 

The sample autocorrelation of the $k$th ($k\geq0$) distance bin is then calculated as 
\begin{align}
\begin{split}
	&\rho(k\Delta d_{bin}) = \frac{1}{(N_{k}-1)\hat{\sigma}^2}\\&\times\sum_{i,j}\mathrm{I}_{k}\Big(X(d_{i})-\hat{\mu}(d_{i}))\Big)\Big(X(d_{j})-\hat{\mu}(d_{j})\Big), \forall \ j>i,
\label{eq:sacf}
\end{split}
\end{align}
with $1\leq i \leq N-1$ and $2\leq j \leq N$. Here, $X(d)$ is the measured pathloss at distance $d$ and and $\hat{\mu}(d)$ is the estimate of the average pathloss at distance $d$, given by the deterministic part of the estimated pathloss model.  $\mathrm{I}_{k}$ is an indicator function given by
\begin{eqnarray}
	\mathrm{I}_k=
	\begin{cases}
	1, & \text{if}\ \ (k-\frac{1}{2})\Delta d_{bin} < |d_{j}-d_{i}| \leq (k+\frac{1}{2})\Delta d_{bin} \\
	0, & \text{otherwise,}
	\end{cases}
\end{eqnarray}
and $N_{k}$ is the number of samples included in the calculation for the $k$th distance bin. In our case, there are no samples between $0<\Delta d_{bin}\leq1/2$. Lastly, $\hat{\sigma}^2$ is the estimated variance of the large scale fading, given by
\begin{eqnarray}
	\hat{\sigma}^2=\frac{1}{N-1}\sum_{l=1}^{N}\Big(X(d_{l})-\mu(d_{l})\Big)^2.
\end{eqnarray}

\begin{figure}[h]
	\centering
	\includegraphics[width=0.4\textwidth]{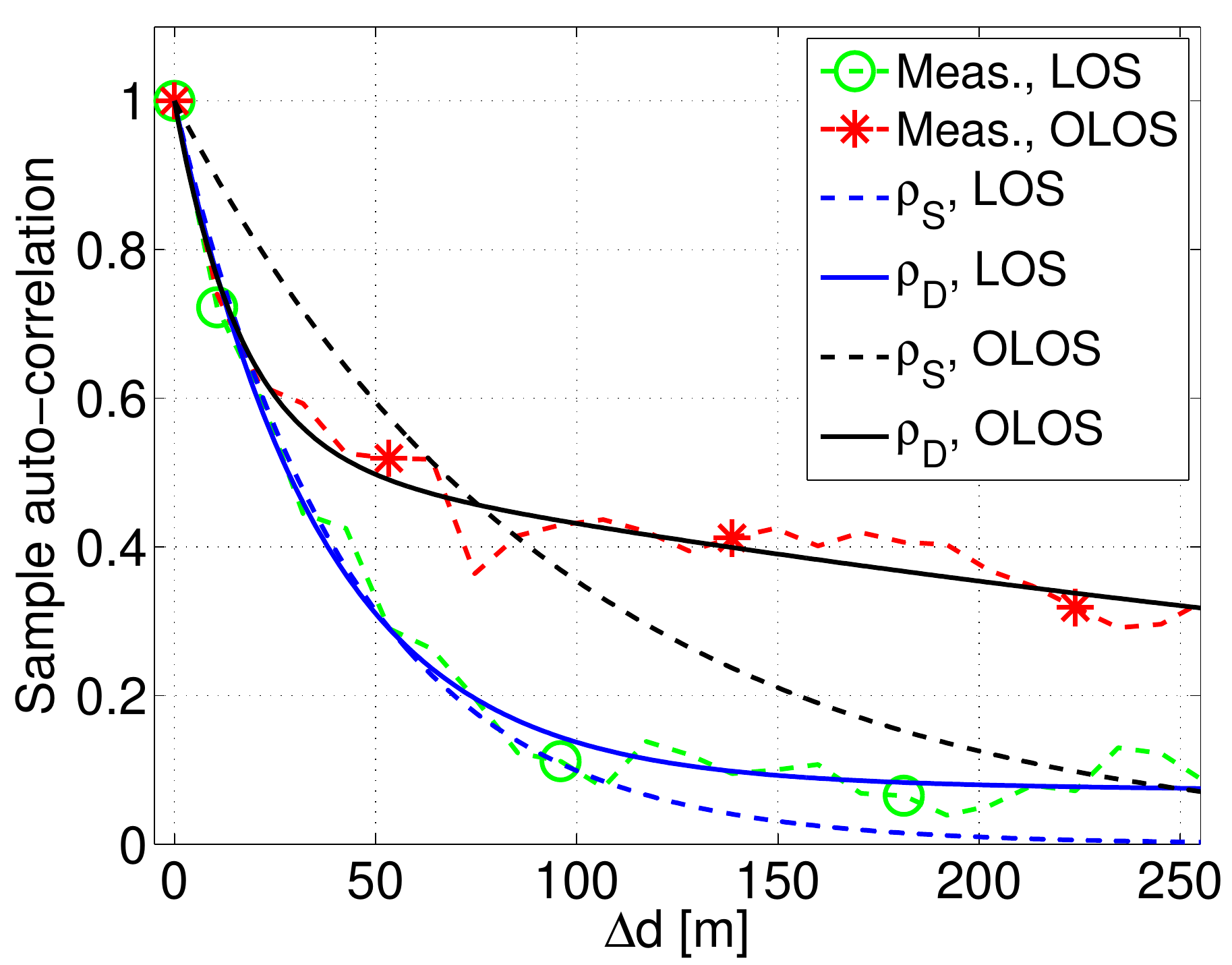}
	\caption{Autocorrelation functions for the large scale fading, both measurements and models according to (\ref{eq:AC_model_1e}) and (\ref{eq:AC_model_2e}) of the communication link XC70 to XC90 in test scenario A, and for both LOS and OLOS cases.} 
	\label{fig:AutoCorr_one}
\end{figure}

The autocorrelation of the shadowing process can be approximated by a well-known model proposed by Gudmundson \cite{Gudmundson91}, based on a negative exponential function,
\begin{equation}
	\rho_{S}(\Delta d)=e^{-\lvert\Delta d\rvert/d_c}.
	\label{eq:AC_model_1e}
\end{equation} 
However, as Fig.\,\ref{fig:AutoCorr_one} shows, the negative exponential model does not fit very well for the OLOS case of the communication link XC70 to XC90 in TS A. This situation has been observed for some other links as well. Therefore, we also model the autocorrelation using a sum of two negative exponential functions,
\begin{equation}
\rho_{D}(\Delta d)=re^{-\lvert\Delta d\rvert/d_{c1}}+(1-r)e^{-\lvert\Delta d\rvert/d_{c2}},
\label{eq:AC_model_2e}
\end{equation} 
which is proposed in \cite{Mawira92}. For (\ref{eq:AC_model_1e}) we focus our attention to modeling the initial decay and therefore values of $0\leq \Delta d\leq100$\,m was used as an input to the ordinary least square estimator of (\ref{eq:AC_model_1e}). For (\ref{eq:AC_model_2e}), values of $0\leq \Delta d\leq500$\,m was used as input to the minimum mean square error estimator. The de-correlation distance, $d_{c}$, is estimated for all communication links in TS A and all links using the middle antenna on the S60 in TS B, and the values are presented in Table\,\ref{tab:values_1e}. Due to space constrains and that the majority of the links using front and rear windscreen antennas in TS B do not fulfill the criteria $d_{max}/d_{min}\geq10$, the de-correlation distance, $d_{c}$ is not presented for these links. The parameters used in (\ref{eq:AC_model_2e}), the de-correlation distances, $d_{c1}$, $d_{c2}$, and the weight factor, $r$, are estimated on the same links as on (\ref{eq:AC_model_1e}), which are presented in the Table\,\ref{tab:values_2e}. The estimated parameters on the two autocorrelation models are both based on the two-ray pathloss model for the LOS case and on the single slope pathloss model for the OLOS case.

\begin{table}[h]
	\renewcommand{\arraystretch}{1.3}
	\caption{Values of the de-correlation distance, $d_{c}$, for the autocorrelation model, $\rho_{S}$.}
	\label{tab:values_1e}
	\centering
	\begin{tabular}{|c|c c|c| c c |}
		\hline
		\textbf{TS A, Link} & \multicolumn{2}{c|}{$d_{c}$ [m]} & \textbf{TS B, Link} & \multicolumn{2}{c|}{$d_{c}$ [m]}\\
		TX-RX & LOS & OLOS & TX-RX & LOS & OLOS \\
		\hline\hline
		
		XC70-S60M & 73.5 & 177.6 & XC70-V70 & 78.0 & \color{red} 299.8 \\ \hline
		
		XC70-XC90 & 43.2 & 96.3 & XC70-XC90 & \color{red} 43.7 & 127.4 \\ \hline
		
		XC70-V70 & \color{red} 38.7 & 89.2 & XC70-S60M &  \color{red} N/A & 170.1 \\ \hline
		
		S60M-XC90 & 68.6 & \color{red} 30.2 & V70-XC90 & \color{red} 59.5 &  \color{red} 38.4 \\ \hline
		
		S60M-V70 & \color{red} 71.8 & \color{red} 83.1 & V70-S60M & \color{red} 76.7 & \color{red} 55.7 \\ \hline
		
		XC90-V70 & \color{red} 60.0 & \color{red} N/A & XC90-S60M & \color{red} 102.2 & N/A \\ \hline
		
		\hline 
	\end{tabular} 
\end{table}

\begin{table}[h]
	\renewcommand{\arraystretch}{1.3}
	\caption{Values of the weight factor, $r$, and the de-correlation distances, $d_{c1}$, $d_{c2}$,  for the autocorrelation model $\rho_{D}$.}
	\label{tab:values_2e}
	\centering
	\begin{tabular}{|c|c c c|c c c|}
		\hline
		\textbf{TS A, Link}  & \multicolumn{3}{c|}{LOS} & \multicolumn{3}{c|}{OLOS} \\
	    TX-RX  & $r$ & $d_{c1}$ [m] & $d_{c2}$ [m] & $r$ & $d_{c1}$ [m] &  $d_{c2}$ [m]\\
		\hline\hline
		
		XC70-S60M & 0.61 & 16.2 & 387.0 & 0.09 &  4.6 & 221.6 \\ \hline
		
		XC70-XC90 & 0.92 & 36.0 & 1953.0 & 0.48 & 16.5 & 511.3\\ \hline
		
		XC70-V70 & \color{red} 0.63 & \color{red} 19.7 & \color{red} 70.4 &  0.57 & 30.1 & 439.8\\ \hline
		
		S60M-XC90 & 0.57 & 11.8 & 315.1 & \color{red} 0.83 &  \color{red} 28.7 & \color{red} 225.7\\ \hline
		
		S60M-V70 & \color{red} 0.54 & \color{red} 14.5 & \color{red} 278.8 & \color{red} 0.58 & \color{red} 17.4 & \color{red} 453.7\\ \hline
		
		XC90-V70 & \color{red} 0.61 & \color{red} 8.7 & \color{red} 202.1 &\color{red} N/A & \color{red} N/A & \color{red} N/A\\
		\hline\hline
		
		\textbf{TS B, Link} & LOS & OLOS & LOS & LOS & OLOS & OLOS\\
		TX-RX  & $r$ & $d_{c1}$ [m] & $d_{c2}$ [m] & $r$ & $d_{c1}$ [m] &  $d_{c2}$ [m]\\ \hline
		
		XC70-V70 & 0.24  & 5.1 & 109.2 &  \color{red} 0.11 & \color{red} 100.4 & \color{red} 326.4\\ \hline
		
		XC70-XC90 & \color{red} N/A & \color{red} N/A & \color{red} N/A & 0.38 & 9.1 & 507.1\\ \hline
		
		XC70-S60M & \color{red} N/A  & \color{red} N/A  & \color{red} N/A  & 0.14 & 4.2 & 248.7 \\ \hline
		
		V70-XC90 & \color{red} 0.53 & \color{red} 12.1 & \color{red} 130.6 &\color{red} 0.36 & \color{red} 43.2 & \color{red} 43.2\\ \hline
		
		V70-S60M & \color{red} 0.30 & \color{red} 4.5 & \color{red} 107.6 & \color{red} 0.70 & \color{red} 18.0 & \color{red} 210.6\\ \hline
		
		XC90-S60M & \color{red} 0.42 & \color{red} 13.8 & \color{red} 294.6 &  N/A &  N/A & N/A\\ \hline
	\end{tabular} 
\end{table}

\begin{figure*}
	\begin{center}
		\subfigure[]{%
		\label{fig:mixModel_Urban}
		\includegraphics[width=.31\textwidth]{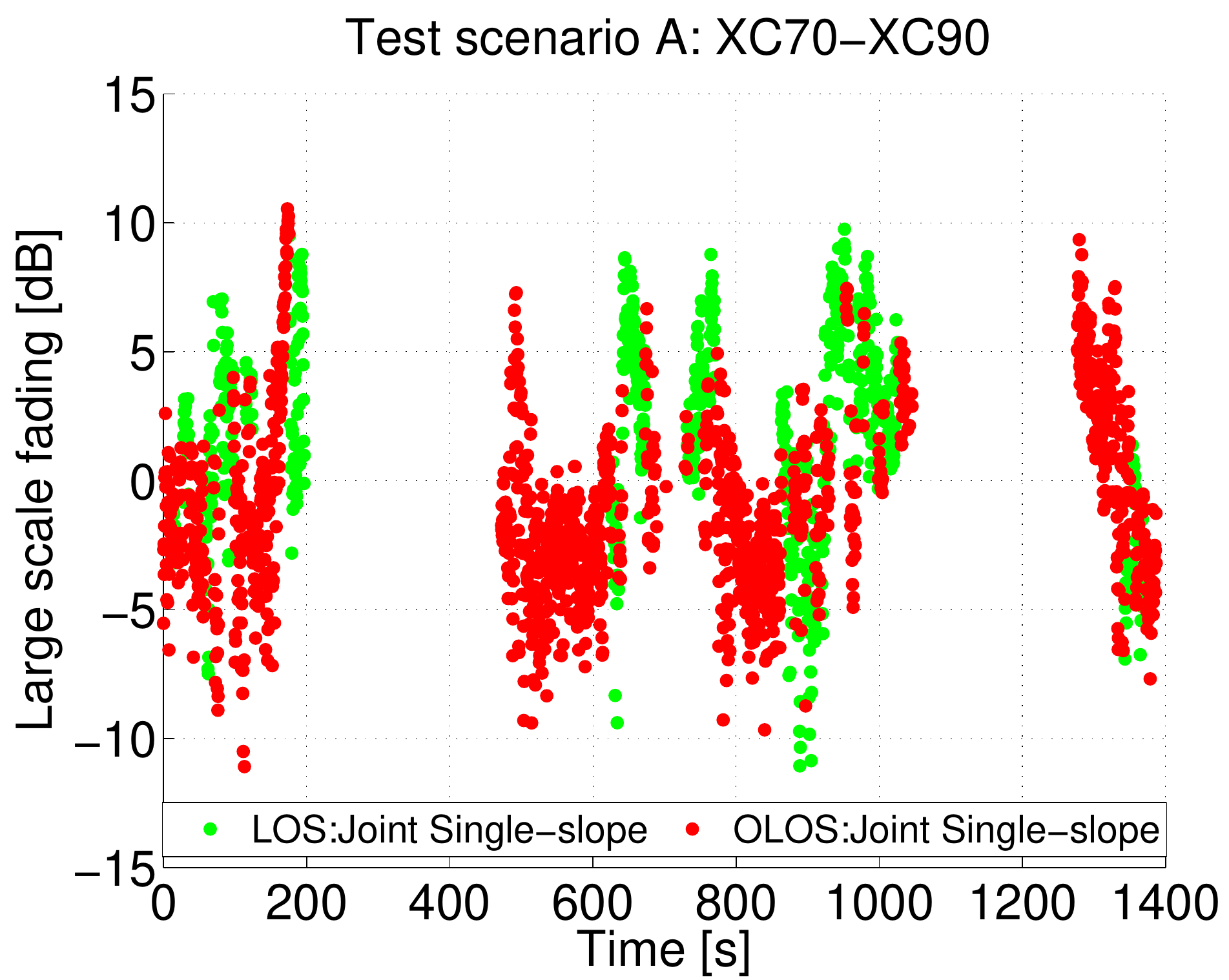}
		}%
		\subfigure[]{%
			\label{fig:pdfLOSvsNLOS_Urban}
			\includegraphics[width=.31\textwidth]{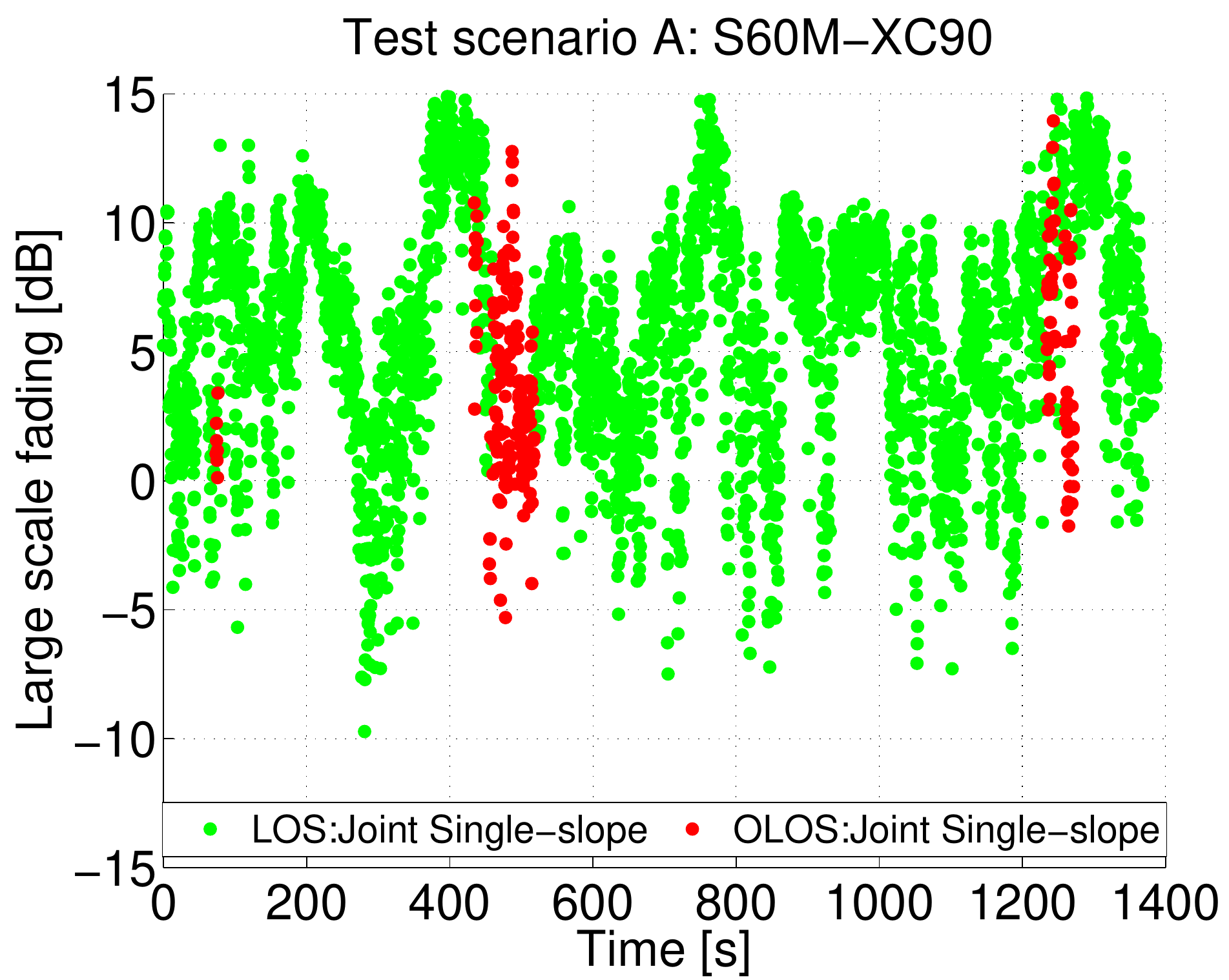}
		}
		\subfigure[]{%
			\label{fig:cdfLOSvsNLOS_Urban} 
			\includegraphics[width=.31\textwidth]{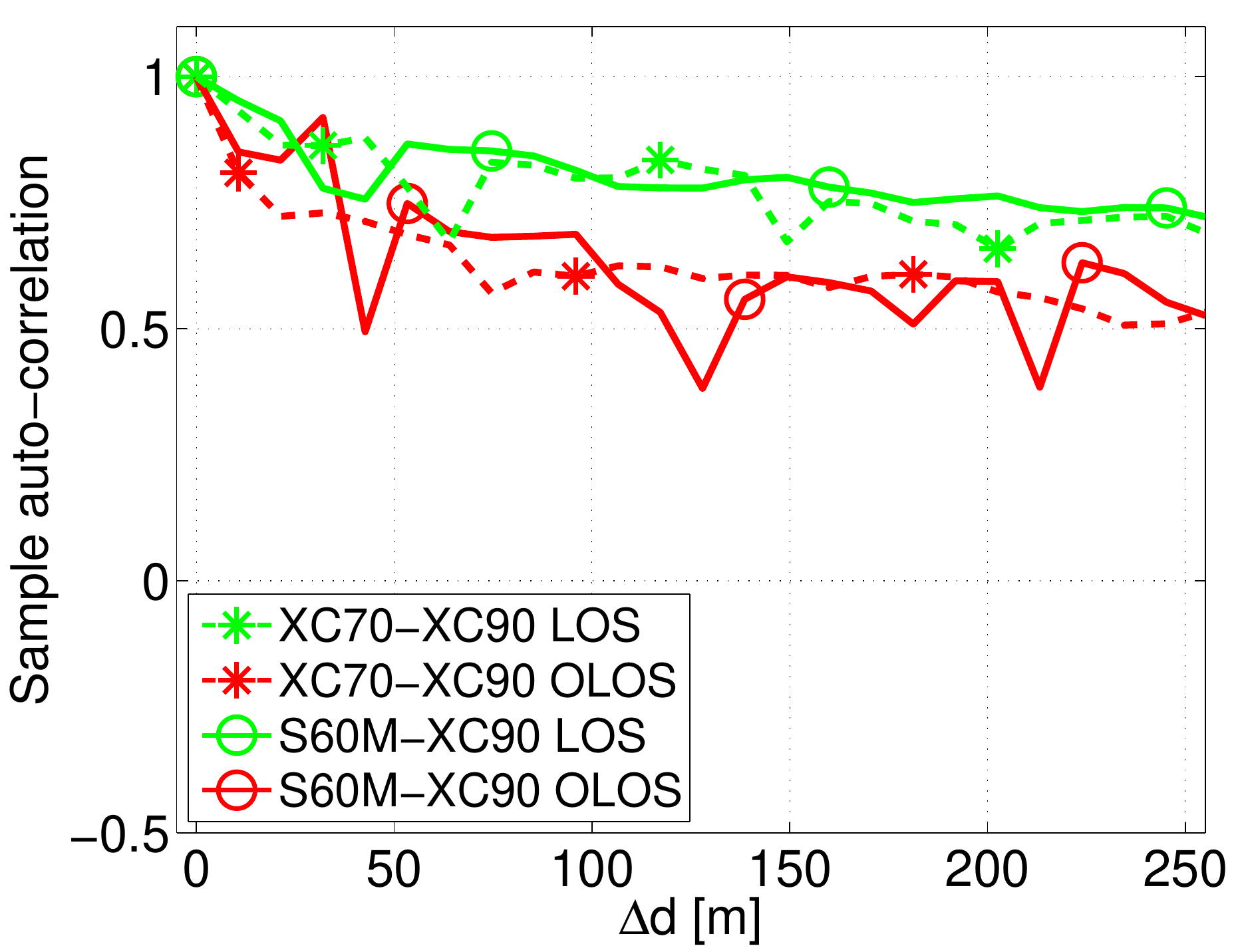}
		}

				
		\subfigure[]{%
			\label{fig:mixModel_Highway}
			\includegraphics[width=.31\textwidth]{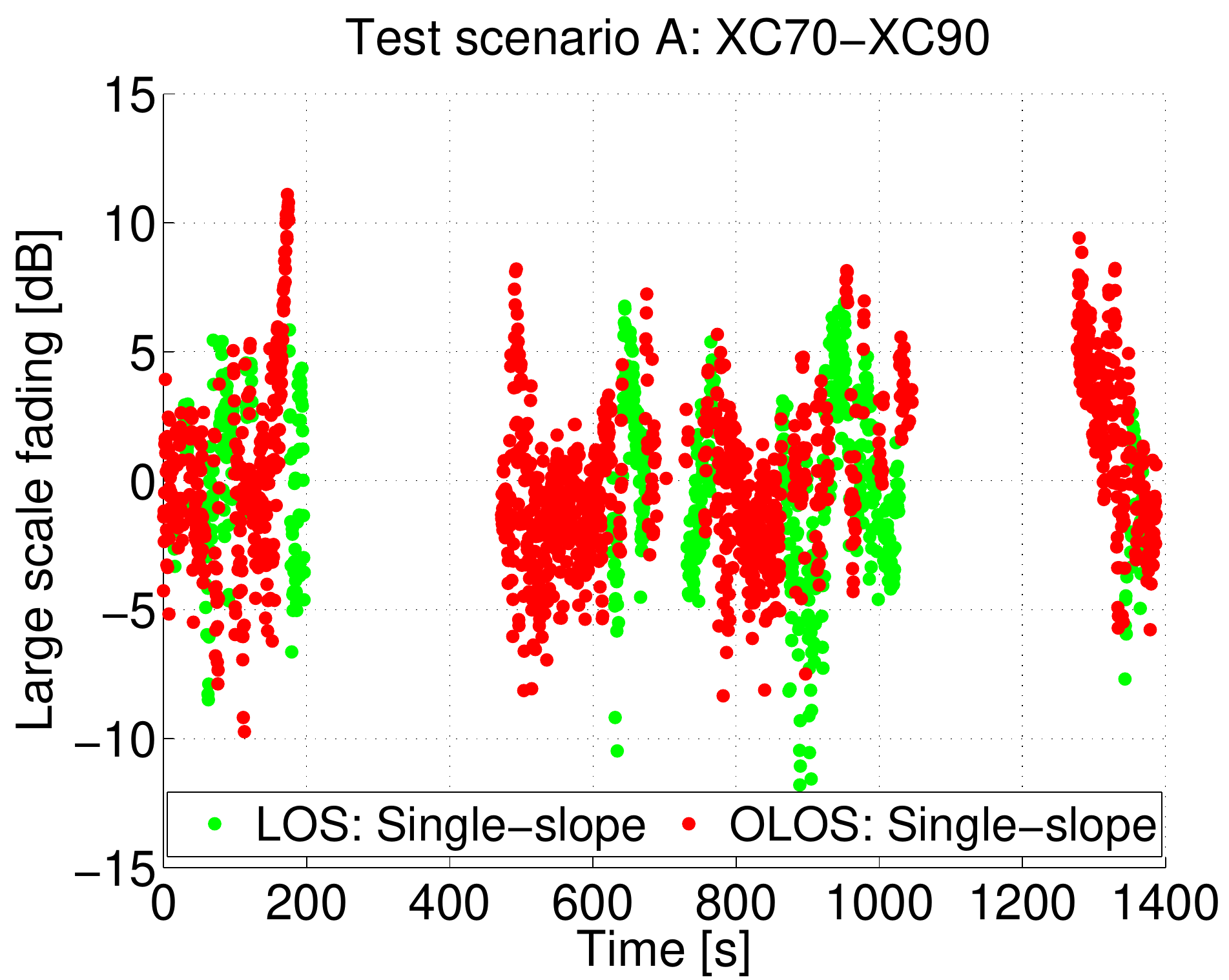}
		}%
		\subfigure[]{%
			\label{fig:pdfLOSvsNLOS_Highway}
			\includegraphics[width=.31\textwidth]{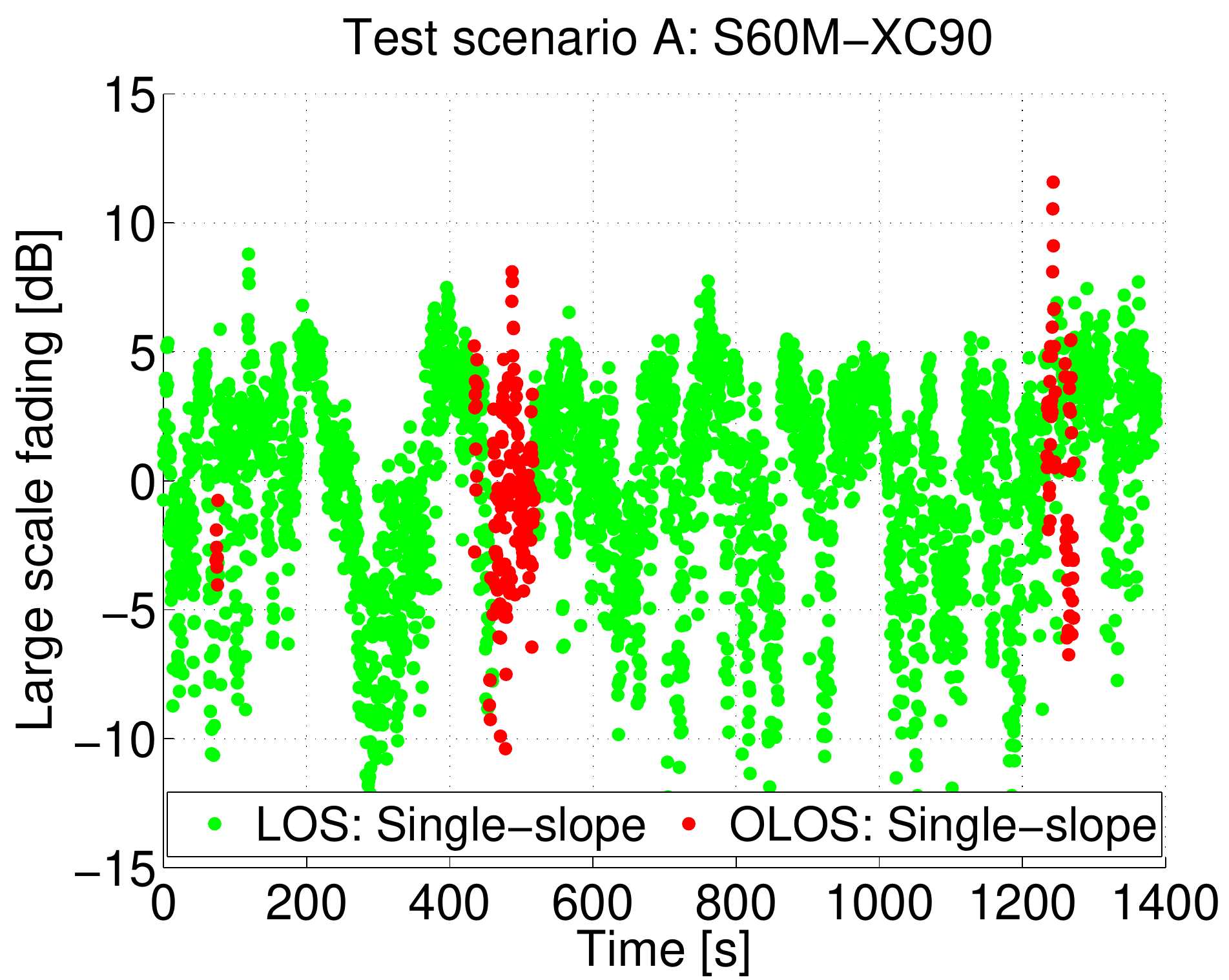}
		}
		\subfigure[]{%
			\label{fig:cdfLOSvsNLOS} 
			\includegraphics[width=.31\textwidth]{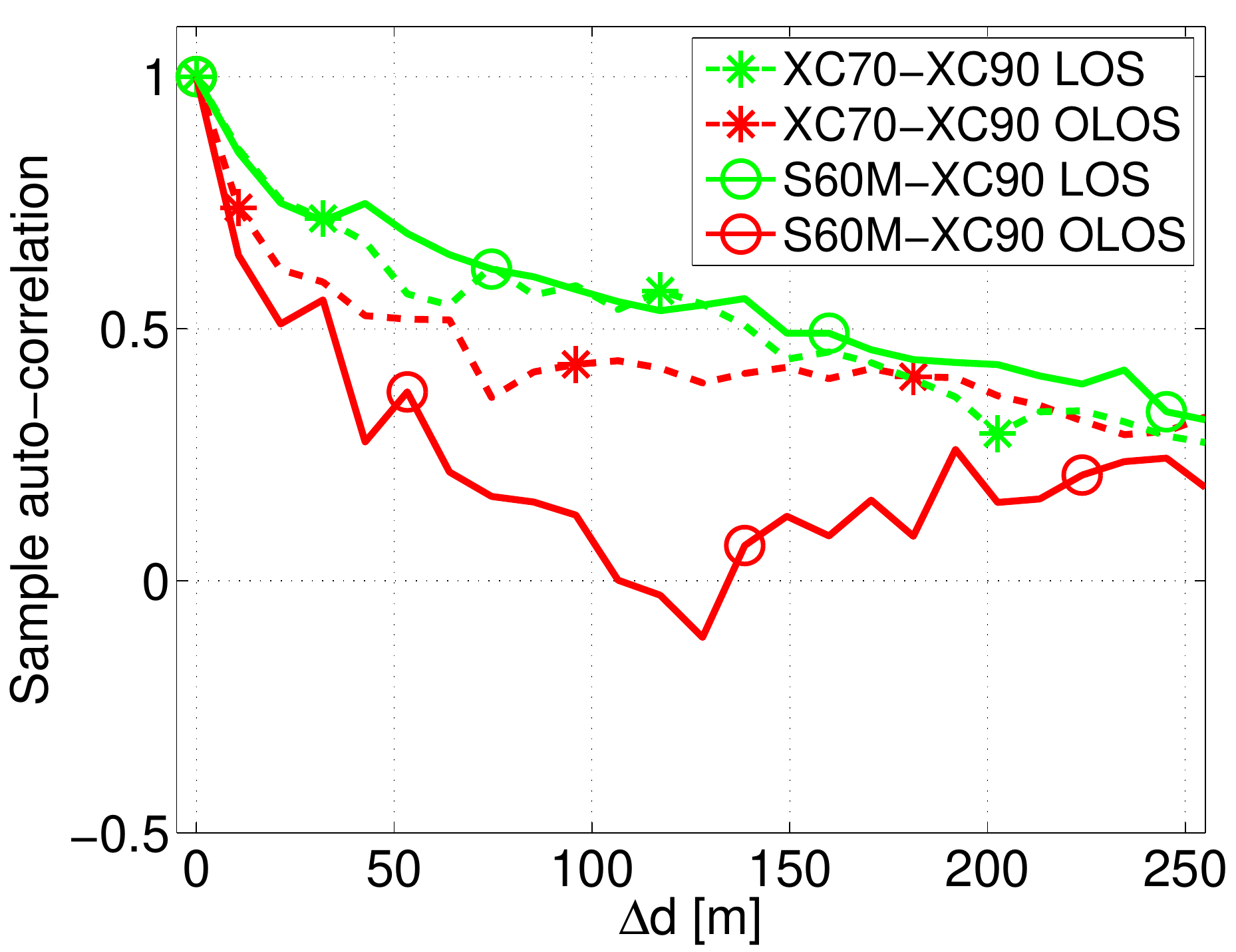}
		}
		
		\subfigure[]{%
			\label{fig:mixModel_Urban}
			\includegraphics[width=.31\textwidth]{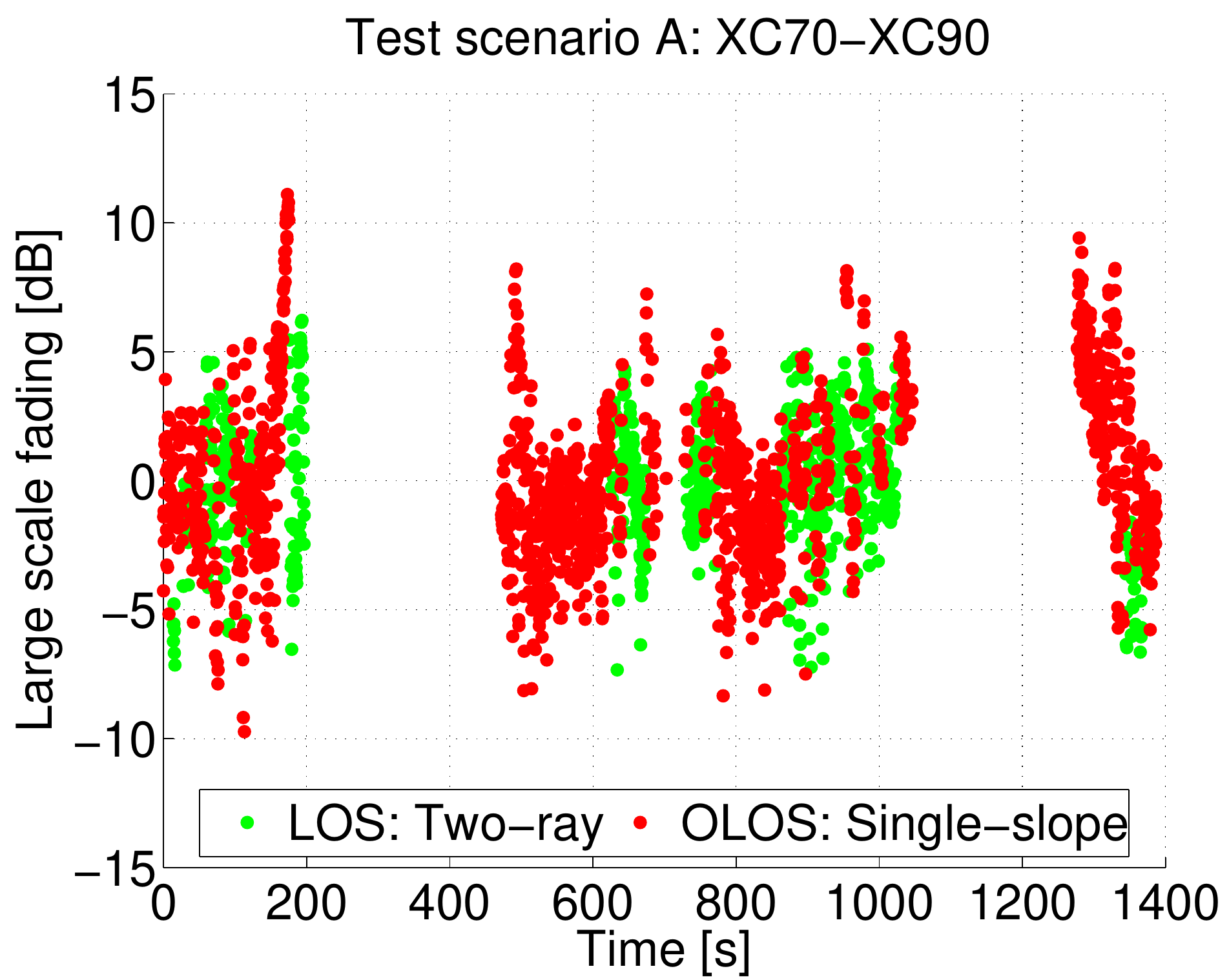}
		}%
		\subfigure[]{%
			\label{fig:pdfLOSvsNLOS_Urban}
			\includegraphics[width=.31\textwidth]{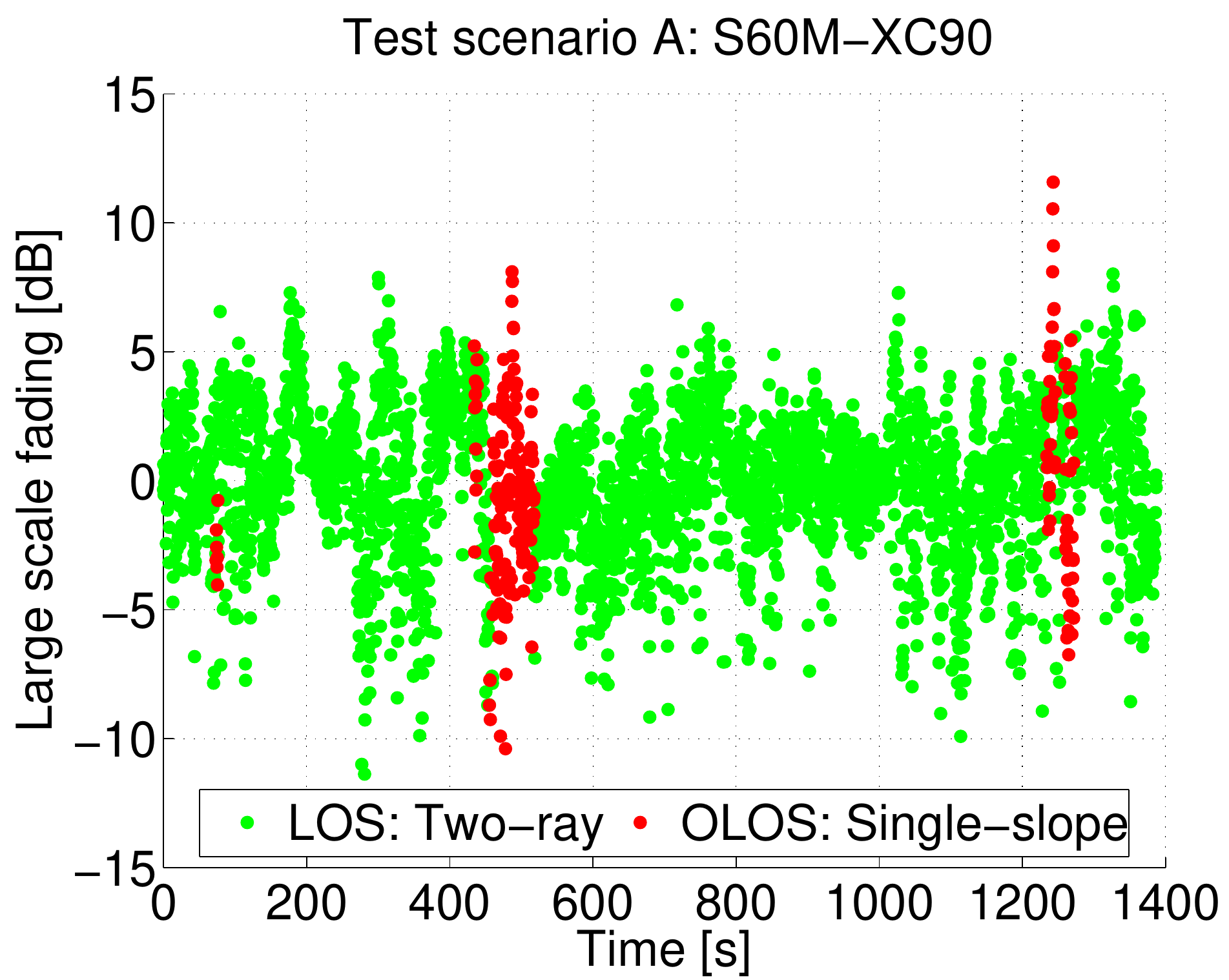}
		}
		\subfigure[]{%
			\label{fig:cdfLOSvsNLOS_Urban} 
			\includegraphics[width=.31\textwidth]{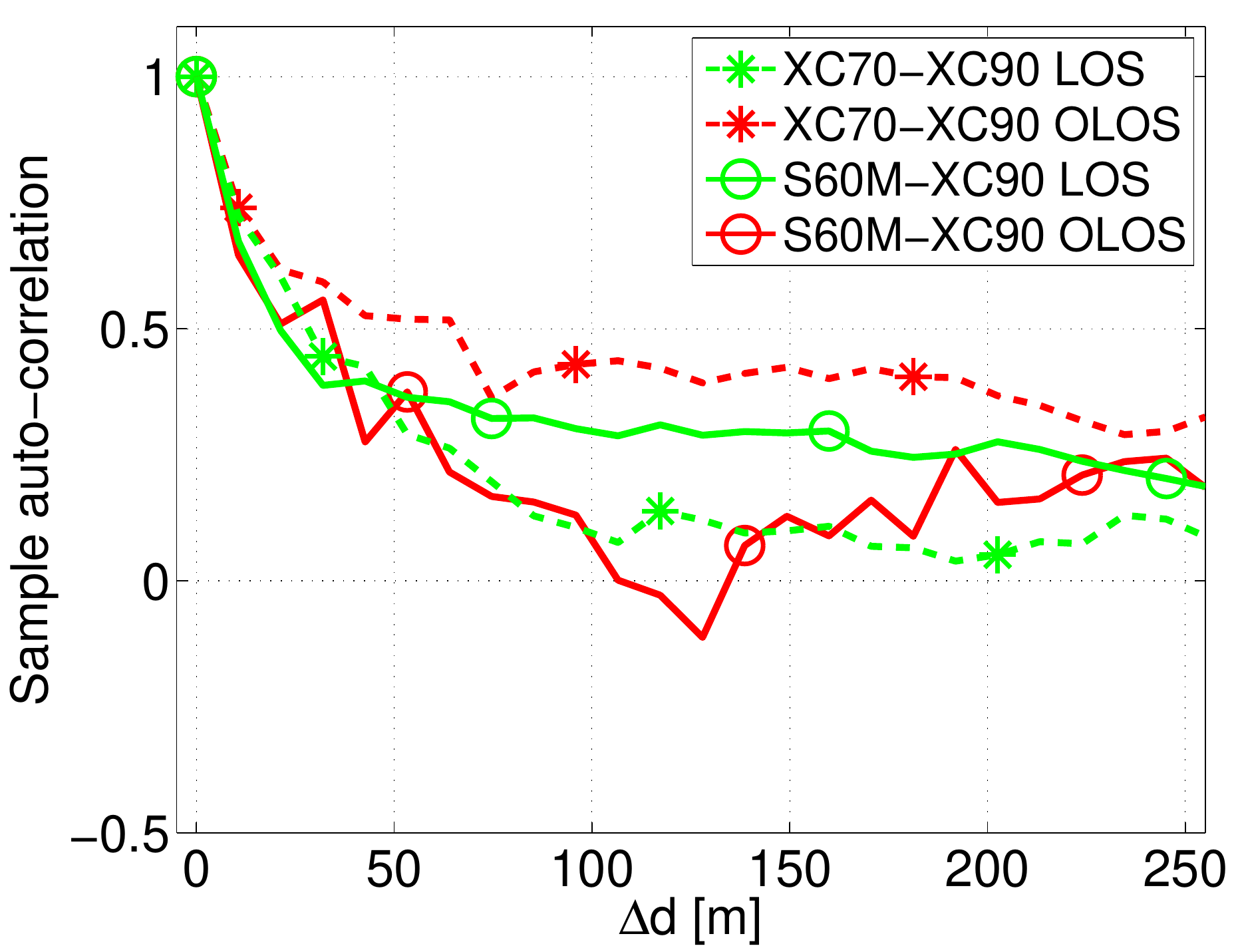}
		}
	\end{center}
	\caption{Figures a, b, c, represent a joint single slope pathloss model for LOS and OLOS of all links together \cite{Nilsson15}, d, e, f, represent a single slope pathloss model for LOS and OLOS separately as well each link individually, when estimating large scale fading and the auto-correlation function, and g, h, i, represent a two-ray pathloss model for LOS and a single slope pathloss model for OLOS, also each link individually. The columns represent: 1) the large scale fading of communication link XC70 to XC90 in TS A. 2) the large scale fading of communication link S60 to XC90 also in TS A. 3) the sample auto-correlation of both links, LOS and OLOS separately.}%
	\label{fig:AutoCorr_diff}
\end{figure*}

An interesting finding regarding analysis of the autocorrelation of the shadow fading process is the different shapes of the curves when using different pathloss models as input to the large scale fading estimation. Fig.\,\ref{fig:AutoCorr_diff} shows the behavior for two different links in TS A. The figure shows clearly the large impact the chosen pathloss model has on the autocorrelation function. A joint single slope pathloss model for LOS and OLOS cases as in \cite{Nilsson15} results in high correlation at large value on $\Delta d$. The reason behind this is that the large scale fading of the two links will have an offset relative to the joint pathloss model due to the differences in the antenna gain patterns for the various links. The joint model is also inaccurate for the LOS case due to its two-ray behavior. Each link is compensated for TX-power and the cable losses in each car of the link, but the antenna gain cannot be compensated due to the fact that the antennas are not omni-directional and that the angle of departure and arrival cannot be estimated in our test setup. Modeling every link individually, as well LOS and OLOS separately, by using a single slope pathloss model, improves the situation. The OLOS becomes less biased or almost zero depending on which of the 18 links that is analyzed but for the LOS case the standard deviation is still large on the large scale fading. Using the two-ray pathloss model for the LOS case improves the situation, the standard deviation on the large scale fading decreases and the correlation reduces dramatically.

\subsection{Multilink Shadowing Correlation}
\begin{figure}[h]
	\centering
	\includegraphics[trim = 1mm 1mm 1mm 1mm, clip=true,width=0.4\textwidth]{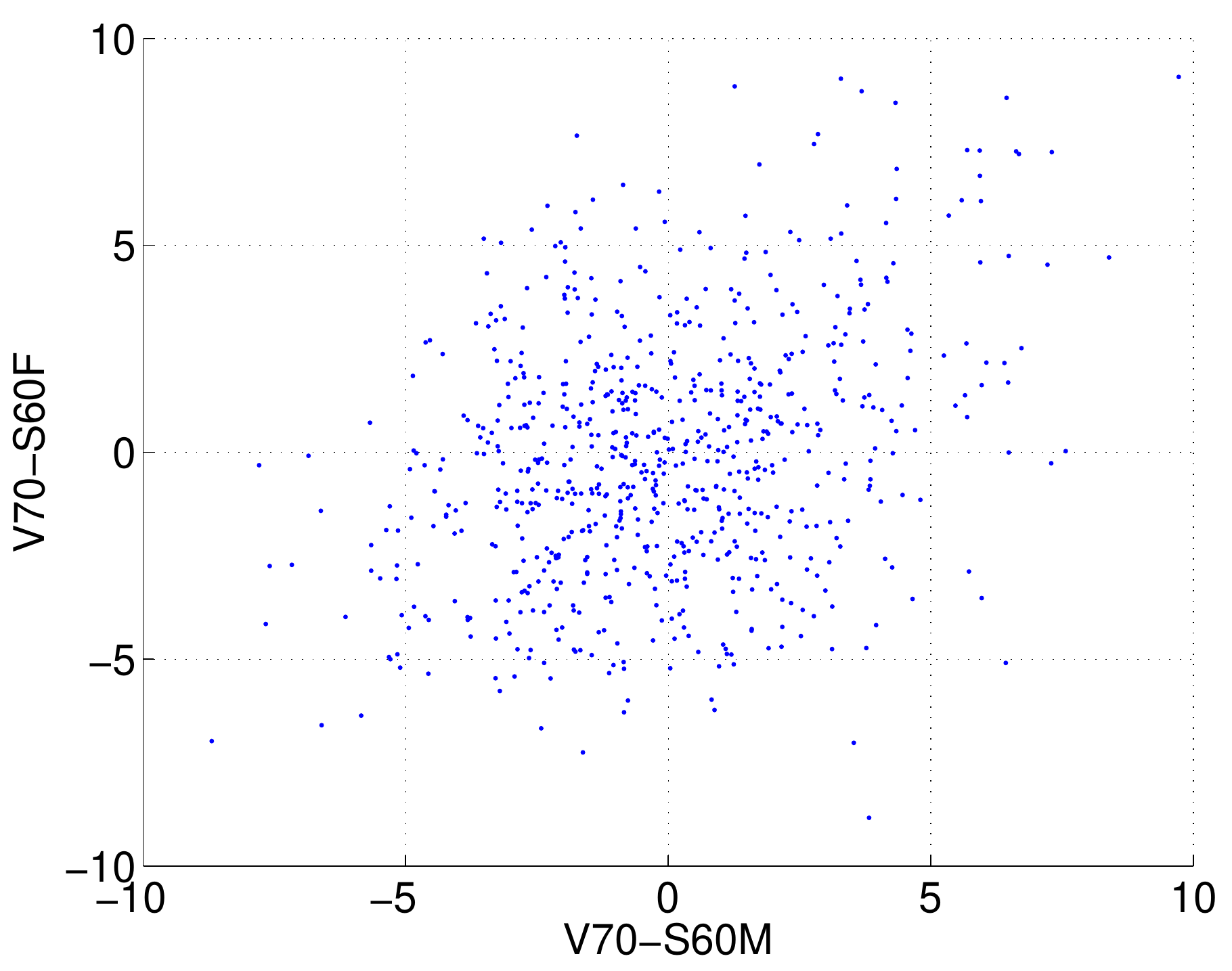}
	\caption{A special case with $\Delta d_{RX}=0$. Scatter plot of the large scale fading for the communication link pairs for the V70 to S60 front antenna (S60F) and for the V70 to the S60 middle antenna (S60M). For this case, the sample cross-correlation, $\rho$, is 0.38.}
	\label{fig:X_corr_same_car}
\end{figure}

Multilink shadowing is a complex mechanism in both time and space, and there might be some non-negligible cross-correlation of the shadowing process when two RX cars receive the same message from one TX car in the VANET. The goal is to find a simple model that can describe the joint shadowing correlation between two communication links by the distance between the two RX cars, $\Delta d_{RX}$, driving on the highway.
In the literature the shadow fading cross-correlation have often been modeled as function of only space, \cite{Szyszkowicz10,Oestges11}, often using a negative exponential model for cellular systems \cite{Wang08,Oestges11,Yamamoto06}, with unit correlation at $\Delta d_{RX}=0$. This kind of negative exponential model is not applicable in our case, due to the differences antenna gain patterns between different cars. If the antenna patterns were very similar, e.g., with dipole antennas located 2 
meters above the car roofs, one would expect a cross-correlation close to one at $\Delta d_{RX}=0$. A special case is with two RX antennas on the same car. The S60 car in our measurement had three different antennas (Front, Middle, and Rear) with very different antenna patterns, see Fig.\,\ref{fig:S60_antenna_pattern}. A TX car was transmitting a signal to two different antennas on the S60. In total, nine different communication links are analyzed regarding the sample cross-correlation (\ref{eq:x_corr}). The $\rho$ in these nine cases are between -0.03 to 0.38 with one outlier at 0.75. For the link pair, V70 to S60 front antenna (S60F) and V70 to the S60 middle antenna (S60M) a scatter plot of the large scale fading is shown in Fig.~\ref{fig:X_corr_same_car}.


To have a large sample size for the estimation of $\rho$, the basis of the measured data is the following four link pairs: 1) TS A, XC70 to XC90 and XC70 to V70. 2) TS A, V70 to S60M and V70 to XC70. 3) TS B, XC70 to XC90 and XC70 to S60. 4) TS B, S60M to V70 and S60M to XC70. The $\Delta d_{RX}$ was grouped into subgroups of 10\,m ($\Delta d_{sub}$=0-10\,m, 10-20\,m, ...). The size of $\Delta d_{RX}$ is chosen to have a large enough sample size within each bin, while maintaining a small enough bin size. For each subgroup the cross-correlation of the large scale fading $(X(d_{i})-\hat{\mu}(d_{i}))$ between all links with the largest distance between TX and RX vs. the large scale fading $(Y(d_{i})-\hat{\mu}(d_{i}))$ links with the shortest distances were estimated using the sample correlation,
\begin{align}
\begin{split}
	&\rho(\Delta d_{sub}) = \frac{1}{(n-1)\hat{\sigma_X}\hat{\sigma_Y}}\\&\times\sum_{i=1}^n\Big(X(d_{i})-\hat{\mu}(d_{i}))\Big)\Big(Y(d_{i})-\hat{\mu}(d_{i})\Big),
	\label{eq:x_corr}
\end{split}
\end{align}
where $n$ is the number of elements within the specific subgroup. As in the estimation of the autocorrelation on large scale fading, $X(d)$ is the measured pathloss at distance $d$ and and $\hat{\mu}(d)$ is the estimate of the average pathloss at distance $d$, given by the deterministic part of the estimated pathloss model. Finally, $\hat{\sigma_X}$ (and the corresponding $\hat{\sigma_Y}$), is the estimated standard deviation of the large scale fading, given by
\begin{eqnarray}
	\hat{\sigma_X}^2=\frac{1}{(n-1)}\sum_{l=1}^{n}\Big(X(d_{l})-\mu(d_{l})\Big)^2.
\end{eqnarray}

As described earlier a negative exponential model is not applicable in our case and therefore a simple linear regression model is used to model $\rho$ as a function of $\Delta d_{RX}$,

\begin{equation}
	\rho_{j1s}(\Delta d_{RX})=
	\begin{cases}
	0.5211-0.0017\Delta d_{RX},&\text{if } \Delta d_{RX}<306\\
	0, & \text{otherwise,}
	\end{cases}
	\label{eq:XC_model_1sset}
\end{equation} 
\begin{equation}
	\rho_{tr,1s}(\Delta d_{RX})=
	\begin{cases}
	0.4674-0.0040\Delta d_{RX},&\text{if } \Delta d_{RX}<116\\
	0,&\text{otherwise}.
	\end{cases}
	\label{eq:XC_model_tr1s}
\end{equation}
Equation (\ref{eq:XC_model_1sset}) is used when using a joint single slope pathloss model for all communication links including LOS and OLOS cases \cite{Nilsson15} when estimating the large scale fading. The equation (\ref{eq:XC_model_tr1s}) is used when each link, with separated LOS and OLOS data, is represented by its own pathloss model when estimating the large scale fading, two-ray for LOS and single slope for OLOS. Only $\Delta d_{RX}$ between 25\,m to 115\,m is used as input to the ordinary least square estimator since small distances are affected to a large degree by the differences of the antenna gain patterns and at large distances the sample size is too small. Hence, the model is only valid up to 120\,m. 
\begin{figure}[h]
	\centering
	\includegraphics[trim = 1mm 1mm 1mm 1mm, clip=true,width=0.4\textwidth]{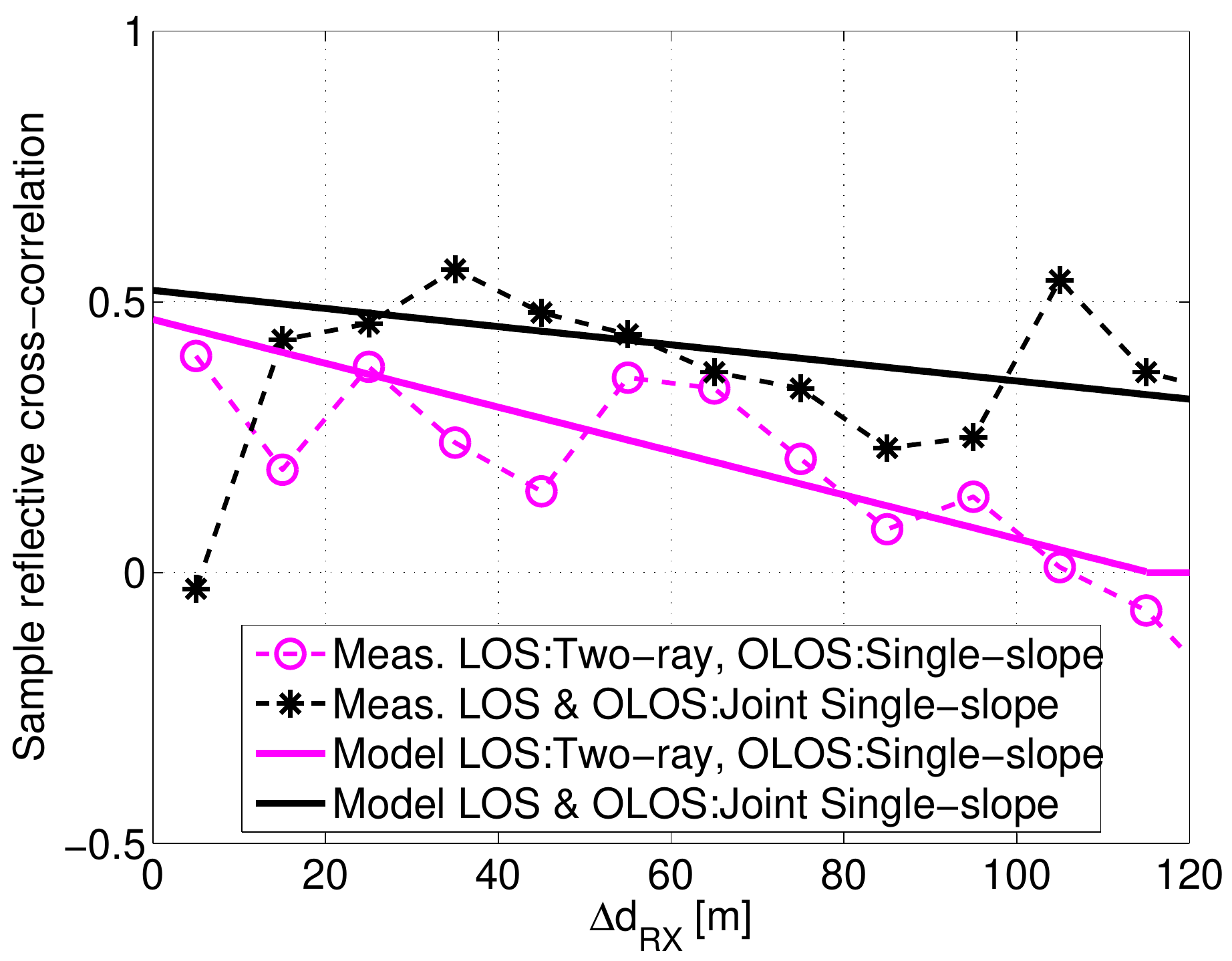}
	\caption{Measured cross-correlation of the large scale fading as a function of the distance between two RX cars, $\Delta d_{RX}$, and the ordinary least square best fit. Two curves are presented: 1) a joint pathloss model \cite{Nilsson15} for all communication links, including LOS and OLOS, are used in the large scale fading estimation 2)  a pathloss model estimated for each link, and for LOS and OLOS separately. }
	\label{fig:X_corr}
\end{figure}

Fig.\,\ref{fig:X_corr} shows the cross-correlation vs. distance between the RX cars for the two cases. The figure also shows the large difference on the slopes of the two curves between the two cases. Even though we are presenting a linear model we are using the de-correlation distance defined in \cite{Gudmundson91}, i.e., $\rho(\Delta d_{RX})=1/e$. Applying the joint single slope pathloss model \cite{Nilsson15} as the basis for the large scale fading on (\ref{eq:XC_model_1sset}) results in a de-correlation distance of 91\,m. Using the two-ray pathloss model for LOS and the single slope model for OLOS, each link separately, on (\ref{eq:XC_model_tr1s}), the de-correlation distance is only 24\,m. This short de-correlation distance means that for many practical highway scenarios the cross-correlation between links can often be neglected with proper modeling and separation of LOS and OLOS situations.


\section{Model Implications}
In this section, we discuss some implications of the modeling approaches and provide some simple examples to show why this is important for VANET simulators. Here, we focus on the effects of the autocorrelation on individual links, and the cross-correlation between different links for the multi-link case. These correlations do not affect the average value of received power if the data ensemble is big enough. However, the correlations will cause the system to experience longer large-scale fading dip durations compared to the uncorrelated case. This is especially important for VANET safety applications, where the consecutive packet error rate is a critical factor. 

\begin{figure}[ht]
	\centering
	\includegraphics[width=0.5\textwidth]{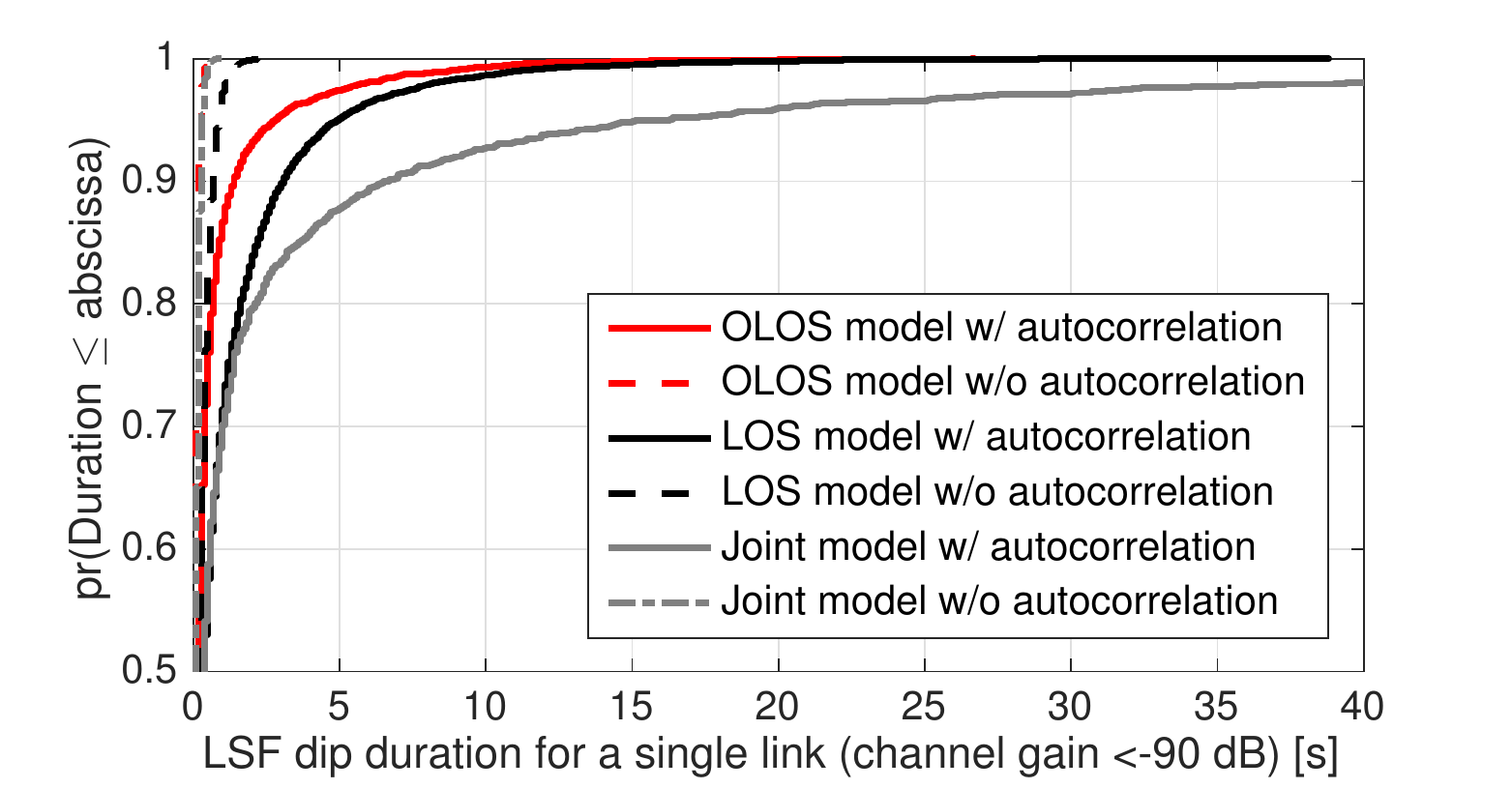}\hspace*{-4mm}\llap{\raisebox{2.7cm}{\includegraphics[trim = 25mm 10mm 123mm 14mm, clip, width=0.28\textwidth]{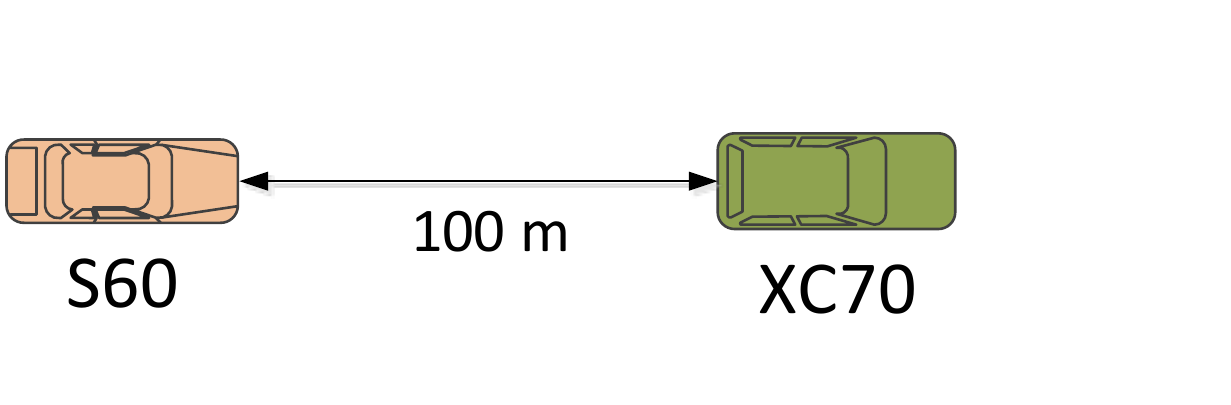}}}\hspace*{4mm}
	\caption{CDF of the modeled large-scale fading dip durations (with channel gains constantly $<$ -90\,dB), for a single link between a XC70 and a S60M. The cars are separated by a distance of 100\,m, and are both driving at a constant speed of 25\,m/s. The LOS (\ref{eq:Two_ray3}) and OLOS (\ref{eq:One_slope}) models are generated using the pathloss parameters in Table II and the autocorrelation model (\ref{eq:AC_model_2e}) using the parameters in Table IV. The joint model is based on the results in \cite{Nilsson15}, with a de-correlation distance of $d_{c}=1500$\,m for the autocorrelation. The same models without any autocorrelation are shown as reference.}
	\label{fig:A_corr1}
\end{figure}

Fig.~\ref{fig:A_corr1} shows the modeled cumulative distribution function (CDF) of the duration of large scale fading dips where the channel gain stays below -90 dB,  for a single link between a XC70 and a S60. Both cars are driving at a constant speed of 25\,m/s and are separated by a distance of 100\,m. The results include the proposed models that distinguishes between LOS and OLOS, with pathloss and autocorrelation parameters from Table II and IV, respectively. The model presented in \cite{Nilsson15}, which uses one joint pathloss for both LOS and OLOS, is also included. Based on the results in \cite{Nilsson15}, we have here assumed an autocorrelation parameter of $d_{c}=1500$ m, using Eq.~\ref{eq:AC_model_1e}. We note that the de-correlation distance, $d_{c}$, for this joint model can vary in the range from about 100 m to 3000 m. The de-correlation distances for the LOS and OLOS models are more consistent and much shorter, as shown in Table III and IV. As a reference, these models are also presented for an autocorrelation being a single $\delta$-function. 

As seen in the figure, the autocorrelation can cause the fading dip durations to sometimes be several seconds long in this specific situation, whereas a model that does not consider any autocorrelation would rarely exhibit fading dips longer than 0.6\,s. It also shows that the joint pathloss model generates much longer fading durations. This is due to the fact that the deterministic behavior of the two-ray model is not being captured by the simple joint pathloss model for all links, which can cause the de-correlation parameter to be artificially large. For some cases, it can be even larger than the value of $d_{c}=1500$\,m that is used in this example \cite{Nilsson15}.  

\begin{figure}[ht]
	\centering
	\includegraphics[width=0.5\textwidth]{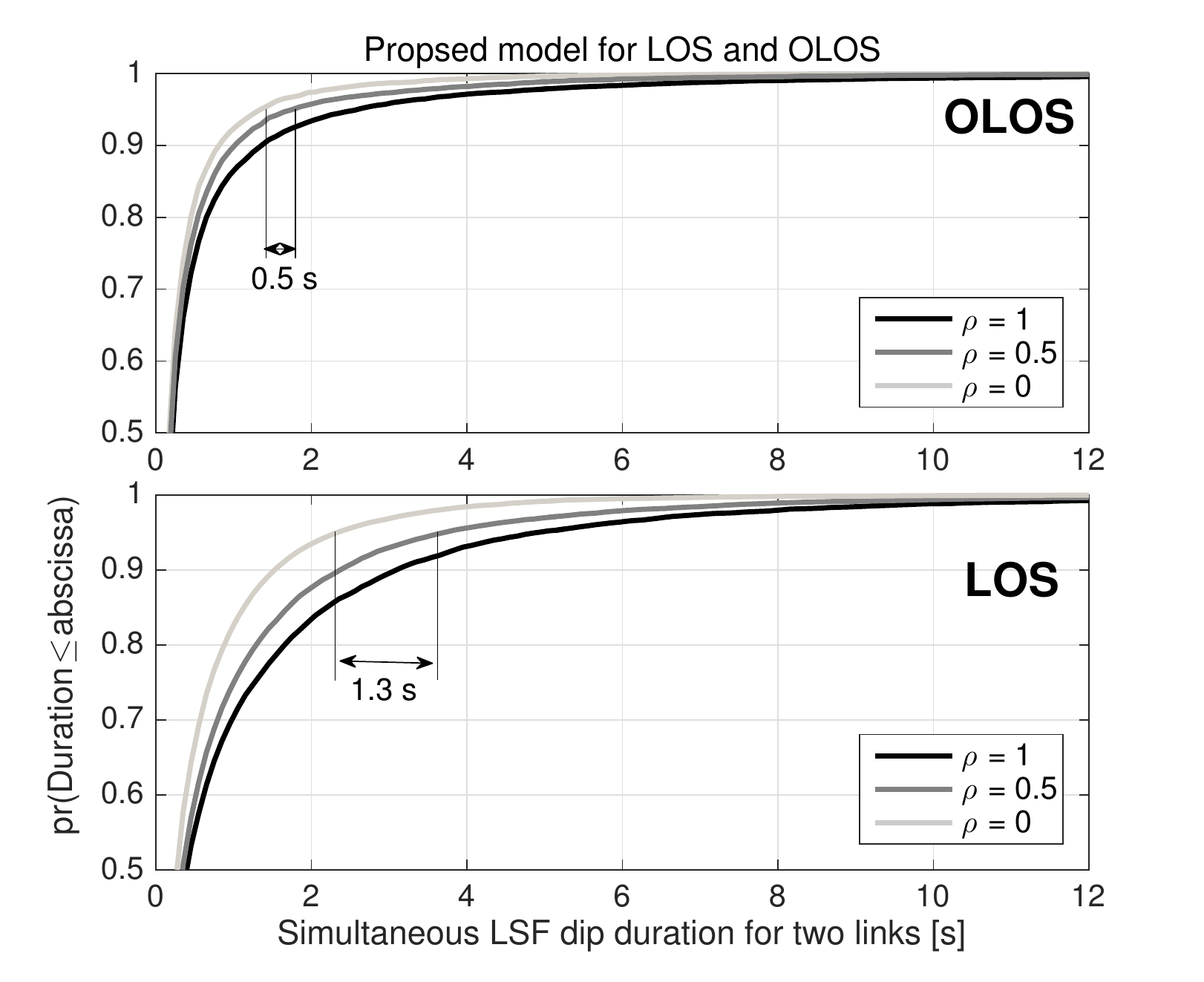}\hspace*{-12mm}\llap{\raisebox{5cm}{\includegraphics[trim = 130mm 10mm 25mm 14mm, clip, width=0.25\textwidth]{1704MN_Simulation_scenario_v1}}}\hspace*{12mm}\hspace*{-12mm}\llap{\raisebox{1.65cm}{\includegraphics[trim = 130mm 10mm 25mm 14mm, clip, width=0.25\textwidth]{1704MN_Simulation_scenario_v1}}}\hspace*{12mm}
	\caption{CDF of the modeled \emph{simultaneous} large-scale fading dip durations (with channel gains $<$ -90 dB), for a XC70 communicating with two different S60 cars at a distance of 100\,m. A special case where $\Delta d_{RX}$=0\,m. The model is based on the proposed models for two LOS links (top) and two OLOS links (bottom), with cross-correlation coefficients $\rho = 0, 0.5$, and 1.}
	\label{fig:X_corrS}
\end{figure}

Fig.~\ref{fig:X_corrS} shows an example similar to Fig.~\ref{fig:A_corr1}, but now for two different links, for a XC70 communicating with two different S60 cars, based on the proposed models that distinguishes between LOS and OLOS. For the two links, we consider cross-correlation coefficients $\rho=0$, $0.5$, and 1. The results show the durations of simultaneous fading dips, where both links experience a channel gain below -90\,dB. This is of importance for relaying techniques, since their performance depend on how often both cars experience severe fading dips at the same time. 
It is observed that the difference between the case with a cross-correlation of $\rho=0.5$ and the uncorrelated case is quite small. The difference is slightly more prominent for the LOS case, which is due to the two-ray fading dip that is located around the Tx-Rx distance of 100\,m. Therefore, the LOS case for this example is a somewhat extreme example. 

In the majority of the cases, the difference between the model with and without cross-correlation is negligible. Furthermore, we note that the impact of the cross-correlation on the simultaneous fading durations is much smaller compared to the variations caused by using the different auto-correlation parameters for different links that are provided in Table III and IV. It is therefore possible to neglect the cross-correlation between different links when using the proposed models for LOS and OLOS for highway scenarios. This is a very useful and practical result, since it makes it much easier to implement VANET simulators for multi-link scenarios. The computational complexity could easily become an issue if the cross-correlation between a large number of links has to be considered.

\section{Conclusion}
Despite the fact that for multilink systems it is essential to model the correlation of the shadowing process for different links, VANET simulations often neglect this cross-correlation, with only a few exceptions. In this paper the importance of separating the measurement data into LOS and OLOS cases is stressed due to the distinct differences in received power as a function of distance between TX and RX. In addition, when estimating pathloss model parameters each link should be estimated separately (not all links together, which is common). An improved two-ray pathloss model for the LOS case to account for differences in antenna gains, car bodies, and antenna pattern is also presented. 

We have demonstrated that the pathloss model applied has a large impact when estimating the correlation functions. It is obvious that when analyzing the de-correlation distance, the basis for the estimation of the large scale fading should be a pathloss model that accurately reflects the large scale fading mean and standard deviation for each link separately. Our findings show that a two-ray model for LOS and a single slope model for OLOS are appropriate ones. These models yield estimates for the de-correlation distance that are smaller compared to a joint pathloss model.

For the multilink shadowing effect we are presenting a simple linear regression model instead of the conventional exponential model used for cellular systems \cite{Wang08,Oestges11,Yamamoto06}, since this conventional model is not applicable in V2V scenarios due to the large differences in antenna gain patterns for different cars. Two models are presented of the cross-correlation for our test setup; one when using a joint pathloss model for all links including LOS and OLOS cases (\ref{eq:XC_model_1sset}), and another one when using a pathloss model for each link, as well LOS and OLOS separately (\ref{eq:XC_model_tr1s}). 
We would like to point out the large difference between the two models, with the joint model having much longer de-correlation distances for both the auto-correlation and the cross-correlation.

The findings regarding pathloss models, autocorrelation behavior and the cross-correlation of the large scale fading processes stress the benefits of geometry based models for VANET simulators. This is also demonstrated by our simple examples of model implications. It is important that the geometry based models distinguish between LOS and OLOS communication and apply different pathloss models for the two cases. Otherwise the VANET simulator needs to consider the cross-correlation between different communication links i.e., implementing (\ref{eq:XC_model_1sset}), to achieve results close to reality. The reason behind this statement is that a de-correlation distance of 91\,m is often larger than typical distances between cars on highways. As a result, the cross-correlation between different communication links needs to be considered in the VANET simulations. For the other case, the de-correlation distance is only 24\,m, and, as shown in Sec. IV, this has a very small impact on the results. Therefore, when using a geometry based model as an input to the VANET simulator, the cross-correlation can be neglected and the implementation of (\ref{eq:XC_model_tr1s}) is not necessary.





\section*{Acknowledgment}

We would like to thank Magnus Eek, Fredrik Huus, Johan Rog\"{o}, Abhijeet Shirolikar, and Dimitrios Vlastaras for their help during the measurements. We also would like to thank Xi Chen fo the master thesis work with the title "Measurement Based Vehicle-to-Vehicle Multi-link Channel modeling and Relaying Performance" \cite{Chen16} were the first steps are taken towards developing a multilink shadowing model.

\ifCLASSOPTIONcaptionsoff
  \newpage
\fi


\bibliographystyle{IEEEtran}
\bibliography{Multilink}


\begin{IEEEbiography}[{\includegraphics[width=1in,height=1.25in,clip,keepaspectratio]{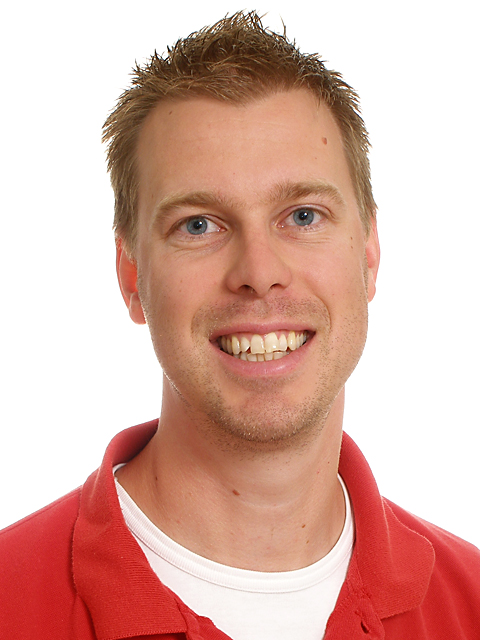}}]{Mikael Nilsson}
	(M'09) received his B.Sc. degree in Electrical Engineering - Radio electronics at V\"axj\"o University, Sweden, in 1997. From 1997 through 2003 he worked mostly as consultant within telecom and space industry. 
	In 2003 he joined Volvo Cars and since 2011 he hold the position as industrial Ph.D. student enrolled at Lund University, Sweden, department of Electrical and Information Technology. His principle research areas are channel characterization of the 5.9\,GHz band and measurement systems, namely the over-the-air mutlti-probe setup for cars. 
\end{IEEEbiography}
\begin{IEEEbiography}[{\includegraphics[width=1in,height=1.25in,clip,keepaspectratio]{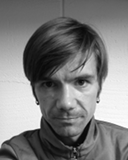}}]{Carl Gustafson}
	 (M'16) received the Ph.D. degree in radio systems and M.Sc. degree in electrical engineering from the Department of Electrical and Information Technology, Lund University, Sweden, where he currently works as a postdoctoral researcher. His main research interests include channel measurements and modelling for mm-wave systems, cellular systems operating above 6 GHz, and vehicular communication systems. Other research interests include massive MIMO, antenna array processing, statistical estimation, and electromagnetic wave propagation. 
\end{IEEEbiography}
\begin{IEEEbiography}[{\includegraphics[width=1in,height=1.25in,clip,keepaspectratio]{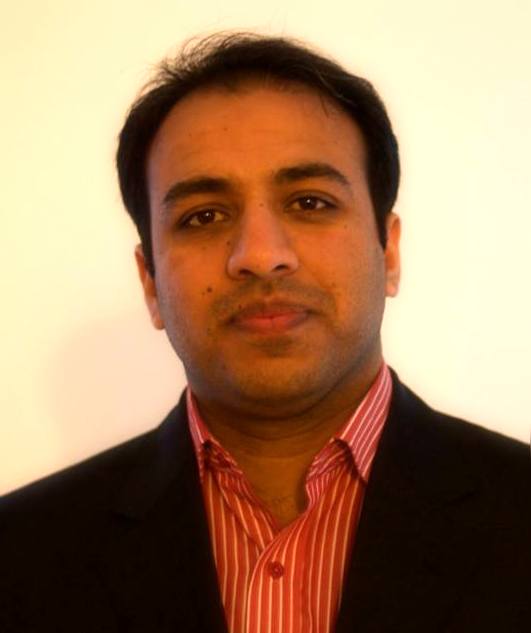}}]{Taimoor Abbas}
	(M'14) received his M.Sc. degree in electronics from Quaid-i-Azam University Islamabad, Pakistan, in 2006, and M.S. degree in wireless communications and Ph.D. degree in radio systems from the Dept. of Electrical and Information Technology, Lund University, Sweden, in 2009 and 2014, respectively. He has been with Ericsson for his master thesis internship. He is now at Volvo Cars. His current research areas include C-ITS, MIMO, 5G systems, and estimation and modeling of radio channels.
\end{IEEEbiography}
\begin{IEEEbiography}[{\includegraphics[width=1in,height=1.25in,clip,keepaspectratio]{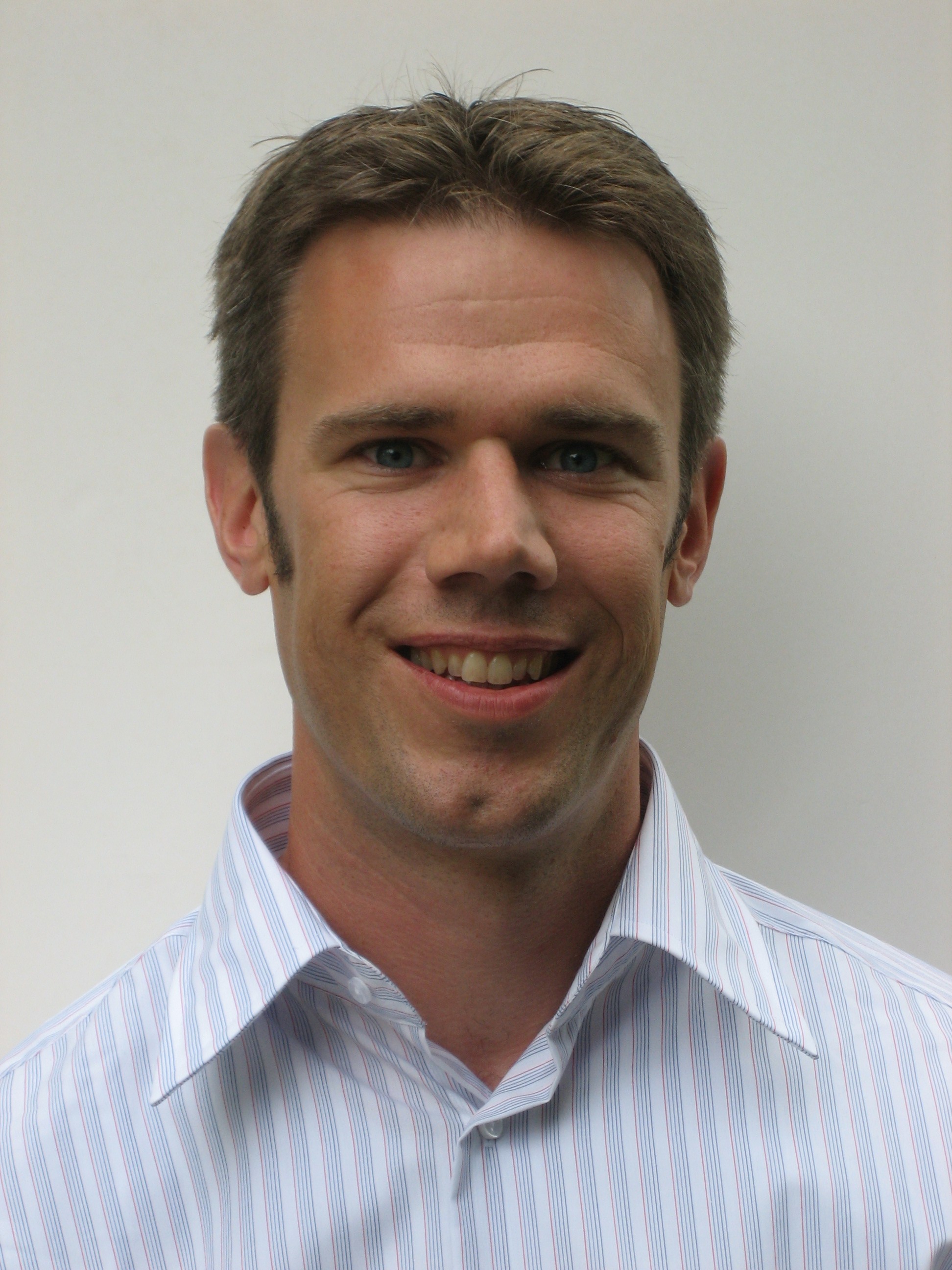}}]{Fredrik Tufvesson}
	(F'17) received his Ph.D. in 2000 from Lund University in Sweden. After two years at a startup company, he joined the department of Electrical and Information Technology at Lund University, where he is now professor of radio systems and is heading the wireless propagation group. His main research interests are channel modeling and characterization for wireless communication, with applications in various areas such as radio based positioning, massive MIMO, UWB, mm wave, and vehicular communication systems.
\end{IEEEbiography}
\end{document}